%% file: iclr2027_conference.tex
\documentclass{article} 
\usepackage{iclr2027_conference,times}

\input{math_commands.tex}

\usepackage{hyperref}
\usepackage{url}
\usepackage[utf8]{inputenc} 
\usepackage[T1]{fontenc}    
\usepackage{hyperref}       
\usepackage{url}            
\usepackage{booktabs}       
\usepackage{amsfonts}       
\usepackage{nicefrac}       
\usepackage{microtype}      
\usepackage{xcolor}         

\usepackage{soul}
\usepackage{amsmath,amssymb,amsthm}
\usepackage{tikz}
\usepackage{braket}
\usepackage{yquant}
\usepackage{tikzscale}
\usepackage{dblfloatfix}
\usepackage{graphicx}
\usepackage{subcaption}
\usepackage{enumitem}
\usepackage{multirow}
\usepackage{array}
\usepackage{pifont}

\usetikzlibrary{chains}
\iclrfinalcopy

\definecolor{drawioBlue}{HTML}{C6DBFA}
\newcommand{\qpmel}{QPMeL}
\newcommand{\qmel}{QMeL}
\newcommand{\pmef}{PMeF}

\newcommand{\C}{\mathbb{C}}
\newcommand{\M}{\mathcal{M}}
\newcommand{\MU}{\mathbb{M}_{\text{unentangled}}}

\newtheorem{definition}{Definition}

\newtheorem{lemma}{Lemma}
\newtheorem{proposition}{Proposition}

\newcommand{\numparagraph}[2]{\refstepcounter{subsection}\label{#1}\paragraph{\thesubsection\ #2}}

\title{\centering Metric Learning with Quantum Feature Spaces}

\author{%
  \end{tabular}
  \hspace{-0.3in}
  \begin{minipage}{\textwidth}
  \centering\normalfont\mdseries
  \begin{tabular}{@{}
    >{\centering\arraybackslash}p{0.33\textwidth}
    >{\centering\arraybackslash}p{0.33\textwidth}
    >{\centering\arraybackslash}p{0.33\textwidth}@{}}
    \textbf{Vinayak Sharma}$^{1}$ &
    \textbf{Lord Sen}$^{2}$ &
    \textbf{Ashish Padhy}$^{2}$ \\
    {\footnotesize\ttfamily Vinayak.Sharma@asu.edu} &
    {\footnotesize\ttfamily 123cs0226@nitrkl.ac.in} &
    {\footnotesize\ttfamily ashishpadhy1729@gmail.com} \\[1.2em]
    \textbf{Sourav Behera}$^{2}$ &
    \textbf{Gennaro De Luca}$^{1}$ &
    \textbf{Shyamapada Mukherjee}$^{2}$ \\
    {\footnotesize\ttfamily souravdipu2002@gmail.com} &
    {\footnotesize\ttfamily Gennaro.Deluca@asu.edu} &
    {\footnotesize\ttfamily mukherjees@nitrkl.ac.in} \\[1.2em]
    \multicolumn{3}{c}{\textbf{Aviral Shrivastava}$^{1}$} \\
    \multicolumn{3}{c}{\footnotesize\ttfamily Aviral.Shrivastava@asu.edu} \\[1em]
    \multicolumn{3}{c}{\small
      $^{1}$Arizona State University, Tempe \quad
      $^{2}$National Institute of Technology, Rourkela}
  \end{tabular}
  \end{minipage}%
  \begin{tabular}{c}
}

\begin{document}

\maketitle

\begin{abstract}
\input{sections/abstract_v2.tex}
\end{abstract}
 
 \section{Introduction} \label{section:Introduction}
 \input{sections/introduction.tex}

\section{Quantum Hilbert Space for Metric Learning} \label{section:analysis}
\input{sections/Analysis_v2.tex}

\section{Quantum Projective Metric Learning (\qpmel{})} \label{section:qpmel}
\input{sections/qpmel_v2.tex}

\section{Results} \label{section:results}
\input{sections/Results_v2.tex}


\section{Conclusion} \label{section:Conclusion}
\input{sections/conclusion_v2.tex}

\bibliography{references}
\bibliographystyle{iclr2027_conference}

\appendix
\numberwithin{equation}{section}
\numberwithin{figure}{section}
\numberwithin{table}{section}

\newpage
\section{Background} \label{section:background}
\input{sections/background_v2.tex}

\section{Extended Background}
\label{appendix:background}
\input{Appendix/Background.tex}

\section{Extended Analysis} \label{appendix:extended_analysis}
\input{Appendix/Analysis_extended.tex}

\section{Proofs and Derivations}
\label{appendix:proofs}
\input{Appendix/proof.tex}

\section{Additional Results} \label{appendix:additional_results}
\input{Appendix/Results_extended.tex}


\end{document}

%% file: math_commands.tex
\usepackage{amsmath,amsfonts,bm}

\def\eqref#1{equation~\ref{#1}}

\def\1{\bm{1}}

\DeclareMathAlphabet{\mathsfit}{\encodingdefault}{\sfdefault}{m}{sl}
\SetMathAlphabet{\mathsfit}{bold}{\encodingdefault}{\sfdefault}{bx}{n}

\newcommand{\E}{\mathbb{E}}

\newcommand{\R}{\mathbb{R}}

\newcommand{\Var}{\mathrm{Var}}



%% file: sections/abstract_v2.tex
The $2^N$ dimensions of Hilbert space give quantum metric learning exponentially many mutually orthogonal states with which to separate classes, but the same exponential dimension also degrades its optimization landscape and the tractability of evaluating its metric. By letting a classical network predict all gate parameters of a fixed ansatz, we isolate the reachable state manifold and its fidelity metric. We prove that this separability is already present on the unentangled manifold $\MU = (\mathbb{CP}^1)^N$, which saturates the $2^N$ maximal orthogonal capacity needed for classification, while entanglement adds states only at intermediate separations, which are costly to resolve under finite sampling. The same size carries costs. Global fidelity causes exponential gradient collapse across both unentangled ($\Theta(c^{-N})$) and entangled ($\Theta(4^{-N})$) manifolds; as fidelities concentrate, the fidelity signal-to-noise ratio at a fixed shot budget also decays exponentially in $N$; and because $\MU$ already holds $2^N$ orthogonal states, expressive, unconstrained encoders under contrastive losses reach trivial minima by \emph{geometric shattering} even without entanglement. We introduce Quantum Projective Metric Learning (\qpmel{}), an exact $\mathcal{O}(N)$ closed-form kernel on $\MU$ that circumvents circuit simulation. \qpmel{} resolves gradient collapse via a curriculum annealing from a $1$-local observable regime (restoring $\Theta(1/N^2)$ polynomial variance) to the global product kernel. Empirically, \qpmel{} as a complete design exceeds eight and matches two of 11 published quantum few-shot designs. On THINGS-EEG2 retrieval, where the published EEG encoders are held fixed and only the quantum head changes, \qpmel{} outperforms explicit variational and re-uploading heads on 200-way zero-shot retrieval, and circuit depth induces kernel concentration. Execution on the 156-qubit IBM Marrakesh processor illustrates the estimation cost that the closed form avoids.

%% file: sections/Introduction.tex
Metric learning trains an encoder so that distance in an embedding space reflects semantic similarity \citep{Chopra2005Metric,TripletOrginal,Schroff_2015}, underpinning few-shot classification \citep{snell2017prototypical} and multimodal retrieval \citep{radford2021learning}. The embedding space matters as much as the encoder: projecting onto a hypersphere bounds distances and stabilizes margin losses \citep{wang2017deep,zhang2020deep}. Quantum metric learning (\qmel{}) replaces this space with the state space of $N$ qubits \citep{lloyd2020quantum,thumwanit2021trainable,kim2022generalization,Liu_2022,hou2023quantum}, motivated by its $2^N$ dimensions \citep{Havl_ek_2019,Schuld_2019} and quantum fidelity $F(x,x') = \bigl|\braket{\psi(x)|\psi(x')}\bigr|^2$, a natural positive-semidefinite kernel \citep{schuld2021supervised}. Metric learning, however, relies on metric geometry: geometric capacity at required margins, optimization landscapes, and evaluation tractability. The exponential dimension serves the first, since $N$ qubits hold $2^N$ mutually orthogonal states with which to separate classes. It also shapes the other two against learning: barren plateaus \citep{mcclean2018barren,cerezo2021cost}, kernel concentration \citep{thanasilp2024exponential}, and the shot cost of resolving concentrated fidelities \citep{miroszewski2024shots} are its consequences, and under instance-level objectives the orthogonal capacity itself can be turned against the embedding. Quantum spaces are nonetheless typically judged by dimension alone. This paper evaluates them as metric-learning targets, weighing the separability their size provides against these costs.

Evaluating quantum state spaces is obstructed in prior models because their circuits act simultaneously as kernel and function approximator \citep{lloyd2020quantum,Liu_2022,hou2023quantum}, preventing failures from being attributed to the geometry or the approximator. Moreover, terminal trainable unitaries cancel identically from fidelity ($F(V\ket{\psi}, V\ket{\phi}) = F(\ket{\psi},\ket{\phi})$), and single-axis encodings confine qubits to 1D great circles. We decouple these roles completely: a classical network $f_w$ predicts every gate parameter of a fixed quantum ansatz $U$ with zero trainable quantum gates, $\ket{\psi_w(x)} = U(f_w(x))\ket{0}^{\otimes N}$ \citep{hur2023neural}. All representation learning is classical; the circuit contributes solely its reachable state manifold and its fidelity metric, isolating the target quantum geometry for rigorous analysis (Section~\ref{subsubsec:dnn_kernel}).

Our primary contributions are:
\begin{itemize}[leftmargin=*,topsep=1pt,itemsep=1pt]
    \item \textbf{Formalization of Geometric Capacity Bounds and Trainability Trade-offs (Lemmas~\ref{lem:packing}--\ref{lem:capacity_variance}):} 
    We prove that the unentangled manifold $\MU = (\mathbb{CP}^1)^N$ saturates the $2^N$ maximal orthogonal capacity needed for classification, while entanglement adds states only at finite separation (Lemma~\ref{lem:packing}), which are costly to resolve under finite sampling (Section~\ref{subsubsec:fidelity}). Furthermore, we show that global fidelity objectives cause exponential gradient collapse across both unentangled ($\Theta(c^{-N})$) and entangled ($\Theta(4^{-N})$) manifolds (Lemma~\ref{lem:capacity_variance}), motivating a trade-off between state capacity and measurement locality.

    \item \textbf{Analysis of Metric Loss Functions under Finite-Shot Fidelity Estimation (FSNR \& Proposition~\ref{prop:shattering}):} 
    We analyze fidelity estimation under finite sampling, showing that, because fidelities concentrate, the fidelity signal-to-noise ratio (FSNR) at a fixed shot budget $S$ decays exponentially with $N$: $\Theta(S \cdot 4^{-N})$ for Haar-random states under the SWAP test and $\Theta(S(2/3)^{N})$ for uniformly spread product states under the inversion test, so holding it fixed costs shots exponential in $N$. Under instance-level contrastive losses, we formalize \emph{geometric shattering} (Proposition~\ref{prop:shattering}): sufficiently expressive, unconstrained encoders exploit the $2^N$ orthogonal capacity of $\MU$ to achieve trivial loss minima without learning semantic representations.

    \item \textbf{\qpmel{} Regime as a Geometric Base Case with Local Training Curriculum (Section~\ref{section:qpmel}):} 
    We introduce Quantum Projective Metric Learning (\qpmel{}), an exact $\mathcal{O}(N)$ closed-form kernel spanning $\MU$ that acts as a minimal geometric base case, training without circuit simulation or shot noise. To resolve the global fidelity barren plateau of Lemma~\ref{lem:capacity_variance}, \qpmel{} trains via an annealing curriculum transitioning from a $1$-local observable regime (which provably restores $\Theta(1/N^2)$ polynomial gradient variance) to the global product kernel.
\end{itemize}

Our empirical results across few-shot classification, multimodal EEG retrieval, and hardware execution are consistent with this analysis. As a complete design (encoder, angle head, kernel and curriculum), \qpmel{} exceeds eight and matches two of eleven published few-shot designs, with fewer or equal qubits in ten. Because these comparisons do not match inputs or encoders, we isolate the quantum head on the much harder multimodal task. On THINGS-EEG2 with the published EEG encoders held fixed, \qpmel{} outperforms explicit variational and re-uploading heads on 200-way zero-shot retrieval ($19.9\%$ against at most $12.3\%$ held-out on the convolutional encoder), and circuit depth primarily induces kernel concentration. Furthermore, execution on the 156-qubit IBM Marrakesh processor illustrates the estimation cost that the closed form on $\MU$ avoids. Finally, contrastive memorization occurs across shallow and deep circuits alike: the $2^N$ orthogonal states available even on $\MU$ let an expressive, unconstrained encoder shatter the batch, and regularizing the encoder, not changing the circuit, prevents it. Together, these findings indicate that the separability Hilbert space offers and the failures of prior quantum metric learners share a source: its exponential dimension provides $2^N$ orthogonal states even on $\MU$, and it also concentrates kernels, flattens the gradients of global fidelity objectives, and lets an expressive encoder shatter. Restricting the reachable manifold to one that keeps that separability, localizing the objective, and regularizing the encoder contain these failures, whereas by the analysis of Section~\ref{sec:hilbert_analysis} reaching further into the state space aggravates them.

%% file: sections/Analysis_v2.tex
\paragraph{Notation.} A quantum embedding maps an input $x$ to an $N$-qubit pure state $\ket{\psi(x)} \in \mathbb{CP}^{2^N-1}$ and compares two inputs by their fidelity $F$, or equivalently their Fubini--Study distance $d_{\mathrm{FS}}$,
\begin{equation} \label{eq:fidelity}
    F(x,x') = \bigl|\braket{\psi(x)|\psi(x')}\bigr|^2 \in [0,1] , \qquad d_{\mathrm{FS}}(x,x') = \arccos\sqrt{F(x,x')} \in [0,\tfrac{\pi}{2}] .
\end{equation}
The states available to the embedding are set by the circuit's dynamical Lie algebra (DLA) $\mathfrak{g}$, the real span of its generators and their nested commutators: the reachable states are the orbit of $\ket{0}^{\otimes N}$ under $e^{\mathfrak{g}}$. Local rotations generate $\bigoplus_q \mathfrak{su}(2)$ of dimension $3N$, with the unentangled manifold $\MU = (\mathbb{CP}^1)^N$ of product states as their orbit, and $F$ factorizes over qubits costing $\mathcal{O}(N)$ to compute (Lemma~\ref{lem:factorized_inner_product}). Generic entangling circuits give $\mathfrak{su}(2^N)$, of dimension $4^N-1$, whose orbit is all of $\mathbb{CP}^{2^N-1}$. Appendix~\ref{section:background} reviews metric-learning objectives, barren plateaus and kernel concentration.

\subsection{Previous quantum metric learning} \label{sec:prior_qmel}
How a circuit arranges its data-dependent, trainable, and fixed gates dictates its reachable states, its fidelity metric, and the degree of classical encoder control. Prior approaches fall into three families:

\subsubsection{Explicit variational quantum circuits (E-VQCs)} \label{subsubsec:enc_vqc}
An E-VQC applies a block $U(\vec{\theta})$ of trainable and fixed gates after a data-dependent encoding $\phi(x)$, $\ket{\Psi_{\theta}(x)} = U(\vec{\theta})\,\phi(x)\ket{0}^{\otimes N}$, as in \citet{chen2024quantum}, \citet{jaderberg2022quantumselfsupervisedlearning} and \citet{wang2024quantum}. This has two consequences:

\textbf{The trainable block cancels from the fidelity.} Any unitary $V$ applied after the encoding, including $V = U(\vec{\theta})$, preserves inner products (Eq.~\ref{eq:terminal_isometry}), so \ul{training $\vec{\theta}$ on a fidelity loss has no effect}.
\begin{equation} \label{eq:terminal_isometry}
    F\bigl(V\ket{\psi(x)},\, V\ket{\psi(x')}\bigr) = \bigl|\bra{\psi(x)} V^\dagger V \ket{\psi(x')}\bigr|^2 = F\bigl(\ket{\psi(x)},\ket{\psi(x')}\bigr) .
\end{equation}
\citet{chen2024quantum} and \citet{jaderberg2022quantumselfsupervisedlearning} instead compare measured expectation values, so their embedding space is the vector of measurement outcomes rather than the state space.

\textbf{The encoder is never given the full manifold.} Several approaches use single-axis angle encoding, such as $R_Y(z_q)$, which limits the encoder to one great circle per qubit, \ul{wasting DNN expressivity}.

\subsubsection{Data re-uploading (DR)} \label{subsubsec:reupload}
DR repeats an encoding followed by a trainable unitary $L$ times, $\ket{\Psi_{\theta}(x)} = \prod_{l=1}^{L} \bigl[U(\theta_l)\,\phi(x)\bigr]\ket{0}^{\otimes N}$, as in \citet{lloyd2020quantum}, \citet{Liu_2022}, \citet{hou2023quantum} and \citet{hammad2023quantum}. Because Eq.~\ref{eq:terminal_isometry} no longer applies, fidelity-based losses can be used. However:

\textbf{Access is restricted to the circuit's spectrum.} The model is a truncated Fourier series in $x$ whose frequencies are fixed by the number of re-uploads \citep{DataReupload,Schuld_2021}, so input placement is bounded by the circuit's spectrum, which limits the access of the classical network $f_w$. \ul{Widening access} with more re-uploads or richer generators \ul{costs trainability} (Section~\ref{bg:bp}).

\textbf{Geometry and approximation cannot be separated.} The circuit both encodes the data and approximates the target, so a \ul{failure is hard to attribute} to the geometry, to insufficient approximation capacity, or to both.


\subsubsection{Classically parameterized quantum kernels (CPQK)} \label{subsubsec:dnn_kernel}
These models treat the circuit as a kernel function with no trainable parameters. For instance, \citet{hur2023neural} and \citet{liu2025neural} train a classical DNN $f_w$ to output the parameters of an entangling $ZZ$ feature map and optimize the fidelity directly. We consider the general form, in which the DNN emits one real parameter for every parameterized gate of a fixed ansatz:
\begin{equation} \label{eq:analytical_setup}
    z = f_w(x) = (z_1,\dots,z_N), \quad z_q \in \R^{L}, \qquad \ket{\psi_w(x)} = U(z)\,\ket{0}^{\otimes N},
\end{equation}
for $N$ qubits and $L$ parameterized gates per qubit, with training on the kernel $k_w(x,x') = |\braket{\psi_w(x)|\psi_w(x')}|^2$. \qpmel{} (Section~\ref{section:qpmel}) is the base case whose access spans $\MU$. In this setup:

\textbf{The setup isolates the geometry.} The circuit defines the accessible quantum states and $f_w$ learns how to map data to them. Therefore, up to approximation error, this setup can represent the mappings of previous work whenever its circuit reaches the same output states (Appendix~\ref{app:upper_bound}).

\textbf{What remains is the geometry.} Because the circuit's parameters are $f_w$'s outputs, the limits of the reachable states act on the encoder directly. A kernel concentrated toward its Haar value, $\E[k_w] = 2^{-N}$ with $\Var[k_w] = \Theta(4^{-N})$ \citep{thanasilp2024exponential}, is a flat loss for $f_w$, and under instance-level objectives the fidelity geometry rewards \emph{geometric shattering} (Proposition~\ref{prop:shattering}), which a sufficiently expressive $f_w$ reaches. \ul{Both apply to any model trained on the fidelity of the states it reaches.}

\subsection{Quantum Hilbert space ($\mathcal{H}_Q$) as a target metric space} \label{sec:hilbert_analysis}
In Eq.~\ref{eq:analytical_setup}, the target metric space is the reachable set $\mathcal{R}(U) = \{\ket{\psi_w(x)}\}$ equipped with fidelity $F$ and distance $d_{\mathrm{FS}}$. An effective embedding space requires: (1)~bounded diameter, (2)~sufficient geometric capacity, and (3)~a trainable, computationally tractable metric. Criterion~(1) is naturally satisfied since $F \in [0,1]$ and $d_{\mathrm{FS}} \leq \pi/2$ for all $N$, and the fixed margin scale it provides can hold exponentially many states at maximal separation, the separability that motivates \qmel{} (Section~\ref{subsubsec:capacity}). The same property admits the trivial minimizers of Proposition~\ref{prop:shattering} (see Appendices~\ref{subsubsec:minimizers}--\ref{subsubsec:smoothness} for trivial minimizers and smoothness). We now examine criteria~(2) and~(3).

\subsubsection{Geometric capacity} \label{subsubsec:capacity}
Capacity refers to how many embeddings a space can keep apart at a given resolution, measured by the maximum cardinality of an $\varepsilon$-separated set.
\begin{definition}[$\varepsilon$-Separated Sets] \label{def:capacity}
A set $\mathcal{S} \subseteq \M$ is $\varepsilon$-separated if $d_{\mathrm{FS}}(\psi,\phi) \geq \varepsilon$ for all distinct $\psi,\phi \in \mathcal{S}$. The capacity $M(\M,\varepsilon)$ is the largest cardinality of such a set.
\end{definition}
A classification loss seeks $K$ classes at maximal separation $\varepsilon = \pi/2$, whereas retrieval tasks require graded similarity preserving a minimum gap $\varepsilon$:

\textbf{Maximal separation.} At $\varepsilon = \pi/2$, $M(\M,\pi/2)$ counts mutually orthogonal states. It is bounded by the dimension of the Hilbert space and attained by the computational basis, \textbf{a set that lies entirely inside $\MU$}:
\begin{equation} \label{eq:orthogonal_capacity}
    M\bigl(\MU, \pi/2\bigr) = M\bigl(\mathbb{CP}^{2^N-1}, \pi/2\bigr) = 2^N ,
\end{equation}
so $N \geq \log_2 K$ qubits suffice: four for the ten classes of MNIST, eleven for the $1623$ characters of Omniglot. Given a function approximator able to realize the assignment, neither entanglement nor further qubits add capacity at this separation. The separability that motivates \qmel{} is therefore a property of the state space itself, not of entanglement, and for the same reason it does not disappear when entanglement is given up. Under an \emph{instance-level} objective such as InfoNCE, \emph{where each sample is its own class}, the same capacity works against the embedding once the encoder is expressive enough to use it (Appendix~\ref{app:shattering}).
\begin{proposition}[Geometric shattering] \label{prop:shattering}
(i) On $\MU$, $F = \prod_q F_q$ is zero if any one qubit pair is orthogonal, and one only if all are aligned. (ii) With fidelity similarity at temperature $\tau$, InfoNCE (Eq.~\ref{eq:infonce}) on $B$ pairs is at least $\log\bigl(1 + (B-1)e^{-1/\tau}\bigr)$, with equality exactly at $k_w(x_i,y_j) = \delta_{ij}$, which $\MU$ attains, by any assignment of samples to orthonormal states, if and only if $B \leq 2^N$.
\end{proposition}
At such a minimum the gradients become data-independent and \ul{are not meaningful outside the training set}; we call this \emph{geometric shattering}. InfoNCE rewards raising a positive and lowering a negative equally, but lowering is easier: one orthogonal qubit sets the fidelity to zero, whereas raising it requires aligning every qubit through multiplicative gradients (Eq.~\ref{eq:mult_gradient}). Under standard cosine similarity, a classical unit sphere cannot reach this floor for $B > 2$ since no three vectors can be pairwise antipodal. On $\MU$, however, the metric reaches its minimum at orthogonality, which $2N$ parameters attain for any $B \leq 2^N$ (as would a classical product kernel on $(S^2)^N$). Shattering therefore needs two ingredients: exponentially many orthogonal states, which $\MU$ already supplies, and an encoder expressive enough to assign a batch to them. Under the same \qpmel{} kernel and InfoNCE objective, a convolutional EEG encoder does not shatter, whereas an unregularized transformer sends a third of its negatives below fidelity $0.01$ (Section~\ref{subsec:trainability_results}). The global measurement is what exposes these states, since a $1$-local kernel admits at most two at zero similarity (Section~\ref{subsec:curriculum}). This is consistent with prior findings that global fidelity kernels on exponentially large spaces drive the Gram matrix toward the identity and memorize, while projected, local kernels do not \citep{huang2021power,thanasilp2024exponential}, although those results concern fixed embeddings rather than trained ones. Because the orthogonal capacity that makes the space useful is present even on $\MU$, giving up entanglement does not prevent shattering, and restricting the circuit enough to remove those states also removes the capacity classification needs (Appendix~\ref{subsubsec:smoothness}). Controlling shattering therefore falls to $f_w$, through a Lipschitz bound \citep{kim2022generalization} or dropout (Section~\ref{subsec:trainability_results}).

\textbf{Finite resolution.} At intermediate $\varepsilon$, capacity is a packing problem. Comparing an $\varepsilon$-separated set with the fraction of the space a Fubini--Study ball (Definition~\ref{def:fs_ball}) occupies gives the bounds:
\begin{lemma}[Capacity at finite resolution] \label{lem:packing}
Let $m = \dim_{\R}(\M)$ be the \textbf{real dimension} of the manifold, with logarithms taken base two so that $\log M$ counts bits. For fixed $\varepsilon \in (0,\pi/2)$,
\begin{equation} \label{eq:packing_bounds}
    m \log_2\frac{1}{\sin\varepsilon} \;\;\leq\;\; \log_2 M(\M,\varepsilon) \;\;\leq\;\; m \log_2\frac{1}{\sin(\varepsilon/2)} + \Delta ,
\end{equation}
with $\Delta = 0$ for $\M = \mathbb{CP}^{2^N-1}$, where $m = 2(2^N-1)$, and $\Delta = \mathcal{O}(N\log N)$ for $\M = \MU$, where $m = 2N$. (Proof in Appendix~\ref{app:packing_argument}.)
\end{lemma}
The log-capacity thus grows with the real dimension of the manifold. At $\varepsilon = \pi/4$, where two states still have fidelity $\tfrac{1}{2}$, the full space holds at least $2^{2^N-1}$ states, while the product manifold holds between $6^N$ and $(4+2\sqrt{2})^{N} N^{N}$ ($2^{\Theta(N\log N)}$ vs.\ $2^{\Theta(2^N)}$). Entanglement therefore adds states at intermediate separations, not at maximal separation, and the closer together those states lie, the more shots it takes to tell them apart: resolving separation $\varepsilon$ costs $S \in \Omega(\varepsilon^{-2})$ shots (scaling to $\Omega(\varepsilon^{-4})$ under SWAP tests for near-orthogonal pairs; Section~\ref{subsubsec:fidelity}). A model can use the added states only as far as its shot budget resolves $\varepsilon$, and under concentration most pairs end up nearly orthogonal.

\textbf{Reachable states.} Every state an embedding uses must be reachable. The reachable set is the orbit of $\ket{0}^{\otimes N}$ under $e^{\mathfrak{g}}$, of dimension at most $\dim\mathfrak{g}$, so occupying $m$ real dimensions requires $\dim\mathfrak{g} \geq m$. Local rotations reach all of $\MU$ with $\dim\mathfrak{g} = 3N$. Covering $\mathbb{CP}^{2^N-1}$ requires an algebra that acts transitively on it, and generic entangling ansätze close to $\mathfrak{su}(2^N)$, for which
\begin{equation} \label{eq:dla_quadratic}
    \dim\mathfrak{g} \;=\; 4^N - 1 \;=\; \Bigl(\frac{m}{2}\Bigr)^{2} + m ,
\end{equation}
quadratic in the dimension reached: all but $m$ of these directions leave the state unchanged (proof in Appendix~\ref{app:closure}).

\textbf{Gradient collapse under global fidelity.} Metric learning evaluates pairwise similarity via fidelity $F(x,x') = \bigl|\braket{\psi(x)|\psi(x')}\bigr|^2$, which measures a global rank-$1$ observable $O_G = \ket{\psi(x')}\bra{\psi(x')}$ across all $N$ qubits ($2^N$ dimensions). As proved by \citet{cerezo2021cost}, \textbf{global cost functions suffer from an exponential barren plateau at random initialization, with or without entanglement}:
\begin{lemma}[Gradient collapse under global fidelity] \label{lem:capacity_variance}
Let an $N$-qubit embedding circuit have parameterization $\theta$, and let $C_G(\theta) = 1 - |\braket{\psi(\theta)|\phi}|^2$ be a global fidelity cost against a product target $\ket{\phi}$. On $\MU$, let the per-qubit parameters be independent and uniformly distributed; on $\mathbb{CP}^{2^N-1}$, let the differentiated gate sit between two independent unitary $2$-designs over $\mathrm{SU}(2^N)$. Then the gradient variance vanishes exponentially with $N$ on both manifolds:
\begin{equation} \label{eq:capacity_variance}
    \Var_{\theta}[\partial_\mu C_G] = \Theta(c^{-N}) \;\; \text{on } \MU , \qquad \Var_{\theta}[\partial_\mu C_G] = \Theta(4^{-N}) \;\; \text{on } \mathbb{CP}^{2^N-1} ,
\end{equation}
where $c = 1/\E[F_q^2] > 1$ depends on the angle distribution and the target: $c = 3$ for Haar-distributed single-qubit states and $c = 8/3$ for single-axis rotations against $\ket{0}^{\otimes N}$. Moreover, the parameter volume with $C_G \le \delta$ vanishes as $\mathcal{O}((1/2)^N)$ \citep{cerezo2021cost}. (Proof in Appendix~\ref{app:capacity}.)
\end{lemma}
Lemma~\ref{lem:capacity_variance} averages over uniformly random parameters, so it describes initialization. When the angles are outputs of $f_w$, the relevant distribution is the one $f_w$ induces, and once positives are aligned ($F_q \approx 1$ on every qubit) the bound no longer applies. This is why \qpmel{} can start on a local cost and end on the global kernel (Section~\ref{subsec:curriculum}).

\textbf{The tradeoff: State capacity vs.\ measurement dimension.} Trainability is governed not merely by the manifold, but by the number of dimensions in which fidelity is measured. Measuring fidelity locally on a $k$-qubit subsystem ($k \ll N$) yields polynomial gradient variance $\Omega(1/\mathrm{poly}(N))$ for shallow ($\mathcal{O}(\log N)$-depth) circuits and broadens the optimization gorge \citep{cerezo2021cost}. However, a local kernel admits at most two states at zero similarity, so the full orthogonal capacity requires the global kernel (Eq.~\ref{eq:capacity_variance}). A local cost can be defined for any embedding, entangled or not \citep{cerezo2021cost}; on $\MU$ the $1$-local cost is additionally computable exactly and classically (Section~\ref{subsec:curriculum}).

\subsubsection{The fidelity inner product} \label{subsubsec:fidelity}
Fidelity is a promising similarity: it is a bounded positive-semidefinite (PSD) kernel and can be estimated on hardware for classically intractable states. Three costs, however, grow with the qubit count $N$.

\textbf{Fidelity is estimated, not computed.} Unless the state is classically tractable, resolving a fidelity gap $\delta$ at intermediate fidelities costs $S \in \Omega(\delta^{-2})$ shots; under the SWAP test on near-orthogonal pairs ($F \approx 0$), resolving separation $\varepsilon$ costs $S \in \Omega(\varepsilon^{-4})$ shots, resolving only $\varepsilon \gtrsim S^{-1/4}$ in that regime. The fidelity signal-to-noise ratio (FSNR) quantifies this:
\begin{equation} \label{eq:fsnr}
    \mathrm{FSNR} \;\equiv\; \frac{\Var_{x,x'}\bigl[F(x,x')\bigr]}{\E_{x,x'}\bigl[\Var(\hat{F} \mid F)\bigr]} ,
\end{equation}
FSNR is $S \cdot 2^{-N}$ for Haar-random states using the inversion test, and $\Theta(S \cdot 4^{-N})$ using the SWAP test. For uniformly spread product states under the inversion test it is $\Theta\bigl(S(2/3)^{N}\bigr)$ ($\Theta(S \cdot 3^{-N})$ under SWAP; Appendix~\ref{app:fsnr_derivations}), so holding it fixed costs shots exponential in $N$ across all cases. Hardware noise also biases the estimate, and this error does not shrink as $S \to \infty$ \citep{Wang_2021}.

\textbf{The gradient is multiplicative.} On the product manifold the kernel factorizes, and so does its derivative,
\begin{equation} \label{eq:mult_gradient}
    F = \prod_{q} F_q \qquad \Longrightarrow \qquad \frac{\partial F}{\partial F_q} = \prod_{q' \neq q} F_{q'} .
\end{equation}
One qubit at $F_{q'} = 0$ zeroes the gradient of every other, and at $F_q \approx 1/2$ the per-qubit gradient is $2^{-(N-1)}$, or $5 \times 10^{-4}$ at $N = 12$. This multiplicative decay is the product-state manifestation of the global fidelity barren plateau \citep{cerezo2021cost}, arising with zero entanglement and constant depth. Section~\ref{section:qpmel} resolves this by transitioning from local to global fidelity.

\textbf{Every pair costs a circuit.} A generic state has no efficient classical description, so each pair in a batch needs its own execution: $5 \times 10^{10}$ circuit executions per step for a unimodal batch of $10^4$ at $10^3$ shots ($10^{11}$ for cross-modal), and a query costs $\mathcal{O}(|\mathrm{DB}|)$ runs. On $\MU$, Lemma~\ref{lem:factorized_inner_product} returns $F$ in $\mathcal{O}(N)$. For small misalignments $F$ is approximately a Gaussian kernel whose bandwidth $f_w$ sets, so shattering resembles a kernel that is too narrow (Appendix~\ref{app:locality}).

\subsection{Design implications} \label{sec:design_impact}
\textbf{Kernel metric learning is the most promising.} The metric is fixed by the geometry of the reachable states, and the expressivity of $f_w$ is cheap by comparison (Eq.~\ref{eq:analytical_setup}).

\textbf{Train on local fidelity before global fidelity.} Global fidelity causes barren plateaus and shot bottlenecks across both unentangled and entangled spaces at random initialization (Lemma~\ref{lem:capacity_variance}). A local surrogate keeps gradients polynomial while the encoder leaves this regime; on $\MU$ the surrogate is exact and classically computable, and annealing to the global kernel recovers its $2^N$ orthogonal states (Section~\ref{subsec:curriculum}).

\textbf{Design the encoder for the geometry.} Concentration flattens the loss, and shattering is an expressive $f_w$ spending the orthogonal states the geometry allows, so the more expressive the encoder, the more it needs a Lipschitz bound or dropout.

%% file: sections/qpmel_v2.tex
\label{sec:qpmel_base}
\qpmel{} instantiates Eq.~\ref{eq:analytical_setup} with the smallest ansatz that reaches all of $\MU$, $U = \bigotimes_q R_Z(\gamma_q)R_Y(2\theta_q)$, so $L = 2$: two angles per qubit. These rotations generate $\bigoplus_q \mathfrak{su}(2)$ of dimension $3N$ (versus $4^N-1$ for generic entangling circuits). Its kernel is closed form (Lemma~\ref{lem:factorized_inner_product}) and training is classical, avoiding variational parameter-shift and shot noise. Since E-VQC and re-uploading reach entangled states, $\MU$ is a submanifold of what they prepare; \qpmel{} serves as the unentangled baseline.

\subsection{Split encoder and angular rescaling} \label{subsec:qpmel_arch}

\begin{figure}[b]
    \centering
    \includegraphics[width=0.85\textwidth]{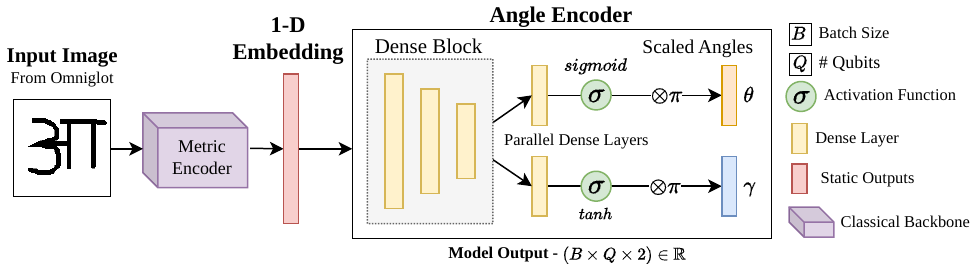}
    \caption{\textit{\qpmel{} architecture. A domain metric encoder maps the input to a latent vector, and an angle head projects it through bounded activations to $N$ coordinate pairs $(\theta_q,\gamma_q)$, one per qubit.}}
    \label{fig:qpmel_arch}
\end{figure}

As shown in Figure~\ref{fig:qpmel_arch}, the encoder is split into a domain model and an angle head. The domain model $h : \mathcal{X} \to \R^d$ is chosen for the modality (a CNN, residual MLP or transformer) and does all of the representation learning. The angle head maps $h(x)$ to $N$ coordinate pairs through bounded activations,
\begin{equation} \label{eq:angle_head}
    \theta_q(x) = \pi \operatorname{sigmoid}\bigl(w_{\theta,q}^{\top} h(x) + b_{\theta,q}\bigr) \in (0,\pi), \qquad
    \gamma_q(x) = \pi \tanh\bigl(w_{\gamma,q}^{\top} h(x) + b_{\gamma,q}\bigr) \in (-\pi,\pi) .
\end{equation}
Because $R_Y$ and $R_Z$ are $2\pi$-periodic, unbounded outputs would map distant inputs to the same state; the bounded activations keep each angle within one period (the chart is still two-to-one). The resulting state is
\begin{equation} \label{eq:qpmel_state}
    \ket{\psi(x)} = \bigotimes_{q=1}^{N} \Bigl( \cos\theta_q(x)\ket{0} + e^{i\gamma_q(x)}\sin\theta_q(x)\ket{1} \Bigr)
\end{equation}

\subsection{The projective metric function} \label{subsec:pmef}

\begin{figure}[t]
    \centering
    \includegraphics[width=0.85\textwidth]{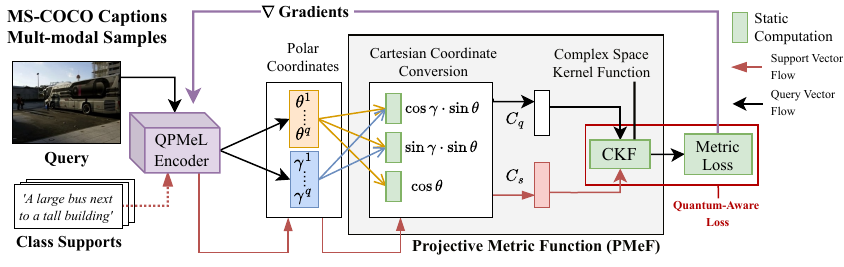}
    \caption{\textit{\qpmel{} training flow, shown for an MS-COCO image and caption. The \pmef{} (shaded) projects the encoder's angles onto real coordinate vectors $C_q$ and computes each qubit's fidelity with the CKF, and the metric loss on the resulting kernel sends gradients back to the encoder. Every block is classical, and no circuit runs during training.}}
    \label{fig:qpmel_training}
\end{figure}

Training never runs the circuit (Figure~\ref{fig:qpmel_training}). Instead it evaluates the fidelity of Eq.~\ref{eq:qpmel_state} classically with the \emph{projective metric function} (\pmef{}), a two-step map from angles to fidelity. The first step maps each qubit's angles to a real unit vector,
\begin{equation} \label{eq:polar_to_cart}
    C_q = (x_q,y_q,z_q) = (\sin\theta_q\cos\gamma_q,\; \sin\theta_q\sin\gamma_q,\; \cos\theta_q) ,
\end{equation}
in which Eq.~\ref{eq:qpmel_state} reads $\bigotimes_q \bigl(z_q\ket{0} + (x_q + iy_q)\ket{1}\bigr)$. The second step computes the fidelity from these vectors alone, since the single-qubit overlap $\braket{\psi_q|\phi_q} = \lambda_q^{r} + i\lambda_q^{c}$ splits into a real and an imaginary part,
\begin{equation} \label{eq:lambda}
    \lambda_q^{r} = x_q x_q' + y_q y_q' + z_q z_q' , \qquad \lambda_q^{c} = x_q y_q' - y_q x_q' .
\end{equation}
\begin{definition}[Complex Kernel Function] \label{def:ckf}
For single-qubit states $\ket{\psi_q},\ket{\phi_q}$ with coordinate vectors $C_q$ and $C_q'$, $\mathrm{CKF}(C_q,C_q') = (\lambda_q^{r})^2 + (\lambda_q^{c})^2 = |\braket{\psi_q|\phi_q}|^2$.
\end{definition}
By Lemma~\ref{lem:factorized_inner_product} the $N$-qubit fidelity is the product of these factors, and composing the two steps gives the model's kernel,
\begin{equation} \label{eq:qpmel_kernel}
    k_w(x,x') = \mathrm{PMeF}(x,x') = \prod_{q=1}^{N} F_q , \qquad F_q = \mathrm{CKF}\bigl(C_q(x), C_q(x')\bigr) ,
\end{equation}
evaluated in $\mathcal{O}(N)$ real arithmetic per pair. Each factor equals $\tfrac{1}{2}(1 + \vec{b}_q \cdot \vec{b}_q')$ for the Bloch vector $\vec{b}_q = (\sin 2\theta_q\cos\gamma_q, \sin 2\theta_q\sin\gamma_q, \cos 2\theta_q)$, an inner product of real features, so the kernel is positive semi-definite by the Schur product theorem (Appendix~\ref{appendix:quantum_kernels}). Training thus avoids circuit simulation, parameter-shift gradients and the shot budget of Section~\ref{subsubsec:fidelity}, and its coordinates prepare the state of Eq.~\ref{eq:qpmel_state} directly.

\subsection{Local observable training and the additive-to-product curriculum} \label{subsec:curriculum}
Section~\ref{sec:hilbert_analysis} established that measuring fidelity globally across all $N$ qubits collapses gradient variance exponentially ($\Theta(c^{-N})$, Lemma~\ref{lem:capacity_variance}), while measuring fidelity in a local observable regime ($k \ll N$) yields polynomial trainability for shallow circuits \citep{cerezo2021cost}. Because states on $\MU$ factorize over qubits, the $1$-local cost surrogate ($k = 1$) is computed exactly and classically: the \emph{additive kernel} $\bar{k}_w$, which averages the single-qubit fidelity factors rather than multiplying them,
\begin{equation} \label{eq:additive_kernel}
    \bar{k}_w(x,x') = \frac{1}{N}\sum_{q=1}^{N} F_q , \qquad \frac{\partial \bar{k}_w}{\partial \theta_q} = \frac{1}{N}\frac{\partial F_q}{\partial \theta_q} .
\end{equation}
This evaluates the $1$-local observable $O_L = \frac{1}{N}\sum_q \ket{\psi_q'}\bra{\psi_q'} \otimes \mathbb{I}_{\bar{q}}$, yielding provably polynomial gradient variance $\Var[\partial \bar{k}_w/\partial \theta_q] = \Theta(1/N^2)$ (Appendix~\ref{app:capacity}) and avoiding the barren plateau of global fidelity. 

However, pure local training has limited margin: $\bar{k}_w$ requires all $N$ qubits to be orthogonal to reach zero similarity, whereas the global product kernel $k_w$ is minimized as soon as a single qubit pair is orthogonal. To navigate the tradeoff between local trainability and global separation advantage, \qpmel{} anneals linearly from the local observable to the global product kernel over $T$ epochs,
\begin{equation} \label{eq:curriculum}
    S_t(x,x') = \alpha_t\, k_w(x,x') + (1-\alpha_t)\, \bar{k}_w(x,x') , \qquad \alpha_t = t/T .
\end{equation}
The two kernels agree on positives ($F=1$ only when all qubits align). The curriculum begins in the local observable regime where gradients are distributed across every qubit independently, and gradually transitions to the global product fidelity, ending on the exact \pmef{} of Eq.~\ref{eq:qpmel_kernel}. Lemma~\ref{lem:capacity_variance} bounds the global kernel's gradients only at random initialization; by the time the curriculum reaches it, positives are aligned and the bound no longer applies.

\noindent \qpmel{} gives the encoder control of all $2N$ real dimensions of $\MU$ and nothing beyond it. A single rotation axis gives the encoder half of these dimensions, and a circuit that reaches beyond $\MU$ can concentrate the kernel or lose input structure (Section~\ref{subsec:trainability_results}); the experiments compare these designs but do not isolate reachable dimension from the other ways they differ.

%% file: sections/Results_v2.tex
\begin{table}[!b]
    \centering
    \footnotesize
    \setlength{\tabcolsep}{3.5pt}
    \caption{\textit{Few-shot classification against published quantum models.}}
    \label{tab:fewshot}
    \begin{tabular}{lllll}
        \toprule
        Model & Circuit & Protocol & Baseline & \textbf{\qpmel{}} \\
        \midrule
        \multicolumn{5}{c}{\textbf{\textit{MNIST}}} \\
        \citet{thumwanit2021trainable} & E-VQC & Binary $(3,6)$ & $91.7$ (9) & $\mathbf{99.80 \pm 0.20}$ (5) \\
        NQE \citep{hur2023neural} & CPQK  & Binary $(0,1)$ & $99.50$ (12) & $99.60 \pm 1.71$ (5) \\
        \citet{Huang} & E-VQC  & 2w-10s $(3,5)$ & $89.69$ (9) & $\mathbf{99.00 \pm 1.01}$ (5) \\
        \citet{wang2024quantum} & E-VQC  & 3w-1s $(0,1,2)$ & $50.5$ (6) & $\mathbf{98.27 \pm 1.86}$ (5) \\
        \citet{wang2024quantum} & E-VQC  & 3w-10s $(0,1,2)$ & $82.5$ (6) & $\mathbf{99.73 \pm 0.37}$ (5) \\
        \multicolumn{2}{l}{No Baseline} & 10w-1s $(0\text{--}9)$ & -- & $96.7 \pm 0.6$ (5) \\
        \midrule
        \multicolumn{5}{c}{\textbf{\textit{Fashion-MNIST}}} \\
        NQE \citep{hur2023neural} & CPQK  & Binary $(0,1)$ & $95.5 \pm 0.41$ (12) & $\mathbf{99.60 \pm 0.95}$ (6) \\
        \citet{Huang} & E-VQC  & 2w-10s $(0,1)$ & $96.92$ (9) & $97.80 \pm 1.53$ (6) \\
        \citet{wang2024quantum} & E-VQC  & 3w-1s $(0,1,2)$ & $62.2$ (6) & $\mathbf{94.00 \pm 2.99}$ (6) \\
        \citet{wang2024quantum} & E-VQC  & 3w-10s $(0,1,2)$ & $73.5$ (6) & $\mathbf{96.67 \pm 1.44}$ (6) \\
        \multicolumn{2}{l}{No Baseline} & 10w-1s $(0\text{--}9)$ & -- & $85.03 \pm 0.38$ (6) \\
        \midrule
        \multicolumn{5}{c}{\textbf{\textit{Omniglot}}} \\
        \citet{Huang} & E-VQC  & 2w-10s $(0,1)$ & $93.83$ (9) & $\mathbf{98.00 \pm 0.85}$ (11) \\
        \citet{Liu_2022} & DR  & 5w-5s & $\mathbf{98.0}$ (64) & $96.16 \pm 0.70$ (11) \\
        \multicolumn{2}{l}{No Baseline} & 10w-5s & -- & $90.02 \pm 0.42$ (11) \\
        \bottomrule
    \end{tabular}
\end{table}

\paragraph{Setup.} \qpmel{} reaches exactly $\MU$ and nothing beyond it (Section~\ref{sec:qpmel_base}). Few-shot classification (Section~\ref{subsec:fewshot_results}) compares \qpmel{} as a complete design with published designs, each with its own inputs, encoder and circuit. Multimodal retrieval (Section~\ref{subsec:retrieval_results}) holds the encoder fixed and changes only the quantum head, testing whether additional state access helps the classical encoder attached to it. MNIST \citep{deng2012mnist} and Fashion-MNIST \citep{xiao2017fashionmnist} use a five-layer CNN with 5 and 6 qubits, Omniglot \citep{omniglot} a three-block CNN with 11, and THINGS-EEG2 \citep{gifford2022large} the EEG encoders of \citet{chen2024quantum} against frozen CLIP image features \citep{radford2021learning} with 10. Few-shot tasks train a class-level objective and classify each query by its nearest class prototype; retrieval tasks train the symmetric InfoNCE objective of CLIP.


\subsection{Few-shot classification} \label{subsec:fewshot_results}
$\MU$ holds $2^N$ mutually orthogonal states (Eq.~\ref{eq:orthogonal_capacity}), so a $K$-class task on $N \geq \log_2 K$ qubits should lose nothing to circuits that reach further. Table~\ref{tab:fewshot} compares \qpmel{} as a complete design (encoder, angle head, kernel and curriculum) with the quantum models of five papers, each on its own protocol. \textit{Circuit} is the family of Section~\ref{sec:prior_qmel}, qubit counts are in parentheses, and bold marks a difference beyond \qpmel{}'s 95\% confidence interval.


\qpmel{} exceeds eight of the eleven baselines, matches two and trails one, with fewer or equal qubits in ten of the eleven comparisons. The largest margins are at one shot, where the strongest models of \citet{wang2024quantum} reach $50.5\%$ on 3-way MNIST and $62.2\%$ on 3-way Fashion-MNIST against $98.3\%$ and $94.0\%$. Every baseline reaches beyond $\MU$ with entangling or trainable layers, yet none of these designs turns that added state access into higher accuracy than \qpmel{}, which classifies all ten MNIST and Fashion-MNIST classes with 5 and 6 qubits, close to the $\lceil \log_2 10 \rceil = 4$ that Eq.~\ref{eq:orthogonal_capacity} requires. The one model it trails, a CNN feeding a 64-qubit re-uploading circuit \citep{Liu_2022}, uses nearly six times as many qubits (Appendix~\ref{app:fewshot_details}). Inputs and encoders differ across these designs, so Table~\ref{tab:fewshot} compares them as wholes; Section~\ref{subsec:retrieval_results} isolates the quantum head on matched encoders.

\begin{table}[h]
    \centering
    \footnotesize
    \setlength{\tabcolsep}{2.5pt}
    \caption{\textit{THINGS-EEG2 200-way top-1 accuracy (\%) with only the head changed; $L$ circuit layers.}}
    \label{tab:eeg}
    \begin{tabular}{llccccccc}
        \toprule
        & & & \multicolumn{2}{c}{QSTConv} & \multicolumn{2}{c}{QNervFormer} & \multicolumn{2}{c}{QNervFormer (dropout)} \\
        \cmidrule(lr){4-5} \cmidrule(lr){6-7} \cmidrule(lr){8-9}
        Head & Similarity & $L$ & Train & Held-out & Train & Held-out & Train & Held-out \\
        \midrule
        \citeauthor{chen2024quantum} & $\langle Z \rangle$ cosine & 4 & -- & $3.2$ & -- & $4.3$ & -- & -- \\
        \midrule
        E-VQC & Fidelity & 4 & $1.5 \pm 0.9$ & $1.2 \pm 0.4$ & $10.1 \pm 2.6$ & $0.2 \pm 0.3$ & $2.8 \pm 0.4$ & $2.3 \pm 1.4$ \\
        DR & Fidelity & 1 & $13.1 \pm 6.5$ & $12.3 \pm 5.2$ & $65.0 \pm 17.9$ & $1.0 \pm 0.9$ & $7.9 \pm 6.2$ & $6.6 \pm 4.6$ \\
        DR & Fidelity & 2 & $2.9 \pm 1.8$ & $3.7 \pm 1.2$ & $43.0 \pm 8.0$ & $0.3 \pm 0.6$ & $2.9 \pm 2.4$ & $2.3 \pm 2.0$ \\
        DR & Fidelity & 4 & $0.6 \pm 0.7$ & $0.7 \pm 0.4$ & $6.5 \pm 1.7$ & $0.5 \pm 0.4$ & $0.9 \pm 1.0$ & $0.9 \pm 0.7$ \\
        \textbf{\qpmel{}} & \pmef{} & -- & $\mathbf{28.0 \pm 3.4}$ & $\mathbf{19.9 \pm 3.1}$ & $\mathbf{96.5 \pm 0.8}$ & $1.1 \pm 0.4$ & $\mathbf{21.2 \pm 8.6}$ & $\mathbf{11.8 \pm 5.6}$ \\
        \bottomrule
    \end{tabular}
\end{table}

\subsection{Multimodal retrieval} \label{subsec:retrieval_results}
\textbf{THINGS-EEG2.} This is the matched comparison, on a much harder task. Table~\ref{tab:eeg} keeps the EEG encoders of \citet{chen2024quantum}, convolutional (QSTConv) and transformer (QNervFormer), bit-identical and changes only the head: their E-VQC with a $\langle Z \rangle$ cosine readout, the same E-VQC and its re-uploading variant (DR) at $L$ layers, both scored by fidelity, and \qpmel{}. \emph{Train} is 200-way retrieval among 200 training images and \emph{held-out} the 200-way zero-shot task on unseen concepts; chance is $0.5\%$. Over 10 subjects and 5 seeds on QSTConv, \qpmel{} reaches $23.96 \pm 5.2\%$ held-out against $0.84 \pm 0.8\%$ for the E-VQC. Our E-VQC matches the equations of \citet{chen2024quantum} but not their accuracy, so we compare heads only on shared backbones (Appendix~\ref{app:eeg_details}).

\subsection{Trainability analysis} \label{subsec:trainability_results}
We test the two CPQK failures of Section~\ref{subsubsec:dnn_kernel}, concentration and shattering (details in Appendix~\ref{app:trainability}).

\textbf{Depth concentrates the kernel.} DR's last CNOT ring cancels (Eq.~\ref{eq:terminal_isometry}), so $L$ layers leave $L-1$ rings in the kernel, and each ring removes input structure: at initialization the kernel's rank correlation $\rho$ with input similarity falls from $0.64$ at one layer to $0.46$, $0.06$ and $0$ at two, four and eight, where its mean and spread match Haar-random states. Held-out accuracy falls with each ring while the first-epoch encoder gradient grows nearly 500-fold from one layer to sixteen, so the kernel fails, not the gradient.

\textbf{\qpmel{} controls all of $\MU$.} Without entangling rings, 1-layer DR reduces to a product kernel on $\MU$ with per-qubit factors $1 - \sin^2 w_q \sin^2((x_q - x'_q)/2)$. Its trainable weights $w_q$ adjust contrast, so it outperforms the E-VQC, whose layers cancel, but trails \qpmel{} ($12.3\%$ vs.\ $19.9\%$ on QSTConv; $6.6\%$ vs.\ $11.8\%$ with branch dropout). Its independent polar angles start the kernel concentrated with a multiplicative gradient (Eq.~\ref{eq:mult_gradient}), and runs linger on the $\ln 1024$ loss plateau for 30--125 epochs. \qpmel{}'s additive start instead gives $1.39$ nats of initial range and its loss falls from epoch~1; ending on the product kernel adds $\sim$5 points of held-out top-1 on QSTConv over a purely additive kernel (Table~\ref{tab:curriculum}).

\textbf{The encoder memorizes; an expressive encoder on the global kernel shatters.} Under \qpmel{}, the unregularized transformer memorizes ($96.5\%$ train vs.\ $1.1\%$ held-out), as it does under an additive kernel that can orthogonalize at most two states (Table~\ref{tab:curriculum}), so memorization originates in the encoder, while the product geometry induces shattering. With a batch of $2^{10} \le 2^N$, memorization barely changes matched fidelity ($0.50 \to 0.56$) but cuts unmatched fidelity fivefold ($0.25 \to 0.05$), with $33\%$ of negatives falling below $0.01$ (the cheap direction of Proposition~\ref{prop:shattering}). Branch dropout on the encoder restores generalization ($0.8\% \to 11.8\%$), and the convolutional QSTConv does not shatter under the same kernel and objective, with $0.2\%$ of negatives below $0.01$ (Appendix~\ref{app:trainability}). Shattering thus requires an expressive, unconstrained encoder and, by Proposition~\ref{prop:shattering}, a batch within $2^N$.
\begin{table}[b]
    \centering
    \footnotesize
    \setlength{\tabcolsep}{4pt}
    \caption{\textit{Few-shot accuracy (\%) on IBM hardware and under a device noise model.}}
    \label{tab:hardware}
    \begin{tabular}{llccccc}
        \toprule
        Dataset & Task & SWAP 2q gates & Shots & Exact & IBM-Prague model & IBM Marrakesh \\
        \midrule
        MNIST & 10w-1s & 73 & 128 & $96.7 \pm 0.6$ & $96.3 \pm 0.8$ & $97.6 \pm 3.4$ \\
        Fashion-MNIST & 10w-1s & 93 & 128 & $85.03 \pm 0.4$ & $80.9 \pm 1.5$ & $74.8 \pm 4.2$ \\
        Omniglot & 5w-3s & 184 & 128 & $93.9 \pm 0.7$ & $93.4 \pm 1.2$ & $84.0 \pm 6.0$ \\
        \bottomrule
    \end{tabular}
\end{table}
\subsection{Hardware execution and noisy simulation} \label{subsec:hardware_results}
On $\MU$ the fidelity is exact in closed form, so hardware execution isolates the estimation costs of Section~\ref{subsubsec:fidelity}. Table~\ref{tab:hardware} runs the MNIST, Fashion-MNIST and Omniglot encoders on the 156-qubit IBM Marrakesh processor with SWAP tests (73--184 routed two-qubit gates), and simulates inversion tests under an IBM-Prague noise model without shot noise (Appendix~\ref{appendix:circuit_discrepancy}). On hardware, MNIST keeps its accuracy within error, while Fashion-MNIST and Omniglot lose about 10 points. The SWAP estimate's variance, $(1-F^2)/S$, exceeds the inversion test's $F(1-F)/S$ by $(1+F)/F$, penalizing near-orthogonal negatives most. The closed form on $\MU$ avoids these estimation costs entirely.

%% file: sections/conclusion_v2.tex
The exponential dimension of the quantum Hilbert space gives metric learning exponentially many orthogonal states with which to separate classes, and it is also the source of its failures. Separating the circuit from the function approximator shows that the unentangled manifold already holds the $2^N$ orthogonal states that classification needs, and that entanglement adds states only at intermediate separations, where concentration makes fidelity costly to estimate. Global fidelity has a barren plateau at random initialization on both manifolds, trainable blocks placed after the encoding cancel from the kernel, and under contrastive losses the same orthogonal capacity lets an expressive, unconstrained encoder shatter the batch. These failures are properties of the state space itself: its dimension concentrates the kernel, flattens the gradients of global fidelity and supplies the orthogonal states that shattering uses. Metric learning in quantum feature spaces therefore depends on keeping the separability that dimension provides while containing its costs, by restricting the reachable manifold to one that, like $\MU$, retains all $2^N$ orthogonal states, localizing the objective and regularizing the encoder, rather than on reaching more of the state space.


%% file: sections/background_v2.tex
\numparagraph{bg:metric}{Metric learning.} A metric-learning model has three parts: an encoder $f_w : \mathcal{X} \to \M$ with parameters $w$, a target space $\M$ with a distance $d$ or similarity $s$, and a loss that pulls similar inputs together and pushes dissimilar ones (negatives) apart by a margin (Appendix~\ref{appendix:metric_learning}). Quantum metric learning takes $\M$ to be the state space of $N$ qubits and $s$ to be the fidelity, which makes the model a quantum kernel method (Appendix~\ref{appendix:quantum_kernels}).

\numparagraph{bg:states}{The quantum state space.} Pure states of $N$ qubits are unit vectors in $\mathcal{H} = (\C^2)^{\otimes N}$, a $2^N$-dimensional Hilbert space; identifying vectors that differ by a global phase gives the complex projective space $\mathbb{CP}^{2^N-1}$. Its similarity and distance are the fidelity $F$ and the Fubini--Study distance $d_{\mathrm{FS}}$ of Eq.~\ref{eq:fidelity}.
Since $F = \operatorname{Tr}[\rho\rho']$ is an inner product of the density matrices $\rho = \ket{\psi}\!\bra{\psi}$, it is a bounded positive-semidefinite (PSD) kernel, and $d_{\mathrm{FS}}$ makes $\mathbb{CP}^{2^N-1}$ a metric space of diameter $\pi/2$. Unentangled (product) states $\bigotimes_q \ket{\psi_q}$ form the manifold $\MU = (\mathbb{CP}^1)^N$. Their overlaps factorize, $\braket{\psi|\phi} = \prod_q \braket{\psi_q|\phi_q}$ (Lemma~\ref{lem:factorized_inner_product}), so their fidelity can be computed classically in $\mathcal{O}(N)$ time.

\paragraph{Encodings and circuits.} Data enters through a feature map $x \mapsto \ket{\psi(x)}$ built from gates. Every single-qubit gate is a Bloch rotation $\exp(-i\theta P/2)$ about a Pauli axis $P$, so two rotations about different axes place a qubit anywhere on $\mathbb{CP}^1$ using two real parameters per qubit, the dense angle encoding of \citet{LaRose2020} (Appendix~\ref{appendix:qml_background}). Local rotations act on each factor independently and never leave $\MU$; reaching the rest of $\mathbb{CP}^{2^N-1}$ requires entangling gates. A parameterized quantum circuit (PQC) alternates rotation layers with entangling blocks, $U(\theta) = \prod_{l} W_l V_l(\theta_l)$, and hybrid \qmel{} models either train $\theta$ after a fixed encoding or re-upload the data between layers \citep{DataReupload}, which turns the model into a truncated Fourier series whose spectrum is set by the encoding schedule rather than by the trained parameters \citep{Schuld_2021}.

\numparagraph{bg:dla}{Reachable states and the dynamical Lie algebra.} A parameterized quantum circuit (PQC) applies gates $\exp(-i\theta_l H_l)$ with Hermitian generators $H_l$. Gates with non-commuting generators do not compose additively, so a deep circuit can also move the state along nested commutators of the generators. Their real span is the dynamical Lie algebra (DLA),
\begin{equation} \label{eq:dla}
    \mathfrak{g} = \mathrm{span}_{\R}\bigl\{\, iH_l, \; [iH_l, iH_m], \; \bigl[iH_l,[iH_m,iH_n]\bigr], \; \dots \,\bigr\} \subseteq \mathfrak{su}(2^N),
\end{equation}
so $\dim\mathfrak{g} \leq 4^N-1$ (Appendix~\ref{appendix:trainability_expressivity}). The states a circuit can reach form the orbit of $\ket{0}^{\otimes N}$ under $e^{\mathfrak{g}}$, a manifold of dimension at most $\dim\mathfrak{g}$. Single-qubit rotations give $\mathfrak{g} = \bigoplus_q \mathfrak{su}(2)$, of dimension $3N$, whose orbit is exactly $\MU$; generic entangling layers give $\mathfrak{su}(2^N)$, of dimension $4^N-1$, whose orbit is all of $\mathbb{CP}^{2^N-1}$.

\numparagraph{bg:bp}{Barren plateaus and kernel concentration.} A cost $C(\theta) = \braket{\psi(\theta)|O|\psi(\theta)}$ has a \emph{barren plateau} when $\E_{\theta}[\partial_\mu C] = 0$ and $\Var_{\theta}[\partial_\mu C] \in \mathcal{O}(b^{-N})$ for some $b > 1$: by Chebyshev's inequality the gradient is then exponentially small with high probability and cannot be resolved with polynomially many shots \citep{mcclean2018barren,cerezo2021cost}. When the generators, observable and input state lie in $\mathfrak{g}$, the variance scales as $1/\dim\mathfrak{g}$ times the $\mathfrak{g}$-purities of the input state and the observable \citep{Ragone_2024}. For global fidelity losses ($O = \ket{\phi}\bra{\phi}$), however, the observable has non-local components up to weight $N$, and \citet{cerezo2021cost} proved that global cost functions incur exponential barren plateaus ($\Theta(c^{-N})$) across both unentangled and entangled spaces (Lemma~\ref{lem:capacity_variance}), while local observables ($k \ll N$) preserve polynomial gradients (Appendix~\ref{appendix:trainability_expressivity}). \emph{Kernel concentration} is a separate failure that depends on how widely inputs are spread over the exponentially large state space and occurs even with no trainable parameters \citep{thanasilp2024exponential}. For Haar-random states $\E[F] = 2^{-N}$ and $\Var[F] = \Theta(4^{-N})$, so the Gram matrix $K_{ij} = F(x_i,x_j)$ approaches the identity, the loss cannot tell which pairs are more similar, and resolving its entries on hardware takes exponentially many shots \citep{miroszewski2024shots,Wang_2021}.

%% file: Appendix/Background.tex
This appendix expands Section~\ref{section:background}. Section~\ref{appendix:metric_learning} states the metric-learning objectives referred to in Section~\ref{bg:metric}, Section~\ref{appendix:qml_background} fixes the quantum notation used throughout, Section~\ref{appendix:quantum_kernels} shows why the hybrid models of Section~\ref{sec:prior_qmel} are kernel methods, and Section~\ref{appendix:trainability_expressivity} collects the trainability and concentration results that Section~\ref{sec:hilbert_analysis} builds on.

\subsection{Metric-Learning Objectives} \label{appendix:metric_learning}
The objectives below are those used in this work and in the quantum metric-learning literature. They differ in how negatives are chosen and in what represents a class, but each defines a margin in $\M$ and each rewards an encoder $f_w$ that pushes negatives apart. The second property is the one the analysis turns on: in a space where nearly all pairs are already far apart, an objective can be satisfied without learning a geometry that transfers.

\paragraph{Triplet loss.} \label{appendix:triplet_loss}
Given an anchor $x_a$, a positive $x_p$ from the same class and a negative $x_n$ from a different class, the triplet loss \citep{TripletOrginal,Schroff_2015,hoffer2018deep}
\begin{equation} \label{eq:triplet_loss}
    \mathcal{L}_{\mathrm{triplet}} = \max\bigl\{0,\; d\bigl(f_w(x_a),f_w(x_p)\bigr) - d\bigl(f_w(x_a),f_w(x_n)\bigr) + m\bigr\}
\end{equation}
enforces a separation of at least $m > 0$ between positive and negative pairs. The hinge is one-sided: once a triplet satisfies the margin it contributes no gradient, however its negative is placed beyond that point.

\paragraph{Prototypical loss.} \label{appendix:prototypical_loss}
In an episode with $K$ classes, the prototype of class $k$ is the mean embedding of its $n_k$ support examples, $\mathbf{c}_k = \frac{1}{n_k}\sum_{i:y_i=k} f_w(x_i)$, and each query $x_q$ is scored against every prototype \citep{snell2017prototypical},
\begin{equation} \label{eq:proto_loss}
    \mathcal{L}_{\mathrm{proto}} = -\log\frac{\exp\bigl(-d(f_w(x_q),\mathbf{c}_{y_q})\bigr)}{\sum_{k=1}^{K}\exp\bigl(-d(f_w(x_q),\mathbf{c}_k)\bigr)} ,
\end{equation}
with $d$ typically the squared Euclidean distance. The negatives are the $K-1$ competing prototypes, so their number is fixed by the episode rather than the batch. The prototype is an arithmetic mean, which presumes that $\M$ is closed under averaging. The quantum state space is not: an average of state vectors is in general neither normalized nor invariant to their global phases. Appendix~\ref{subsubsec:minimizers} gives the fidelity-optimal replacement.

\paragraph{Contrastive (InfoNCE) loss.} \label{appendix:infonce}
For a batch of $B$ paired samples $\{(x_i,y_i)\}$, InfoNCE \citep{oord2018cpc}, the objective of CLIP-style multi-modal training \citep{radford2021learning}, is
\begin{equation} \label{eq:infonce}
    \mathcal{L}_{\mathrm{InfoNCE}} = -\frac{1}{B}\sum_{i=1}^{B}\log\frac{\exp\bigl(s(f_w(x_i),g_w(y_i))/\tau\bigr)}{\sum_{j=1}^{B}\exp\bigl(s(f_w(x_i),g_w(y_j))/\tau\bigr)} ,
\end{equation}
with encoders $f_w$ and $g_w$ (shared in the uni-modal case), a bounded similarity $s$ and a temperature $\tau$. It differs from the hinge in two ways that matter here. It has no margin at which the gradient vanishes, so it keeps rewarding a wider gap between the positive and the $B-1$ in-batch negatives; and its number of negatives grows with the batch, so it is dominated by negative pairs.

\paragraph{Proxy and angular-margin losses.} \label{appendix:angular_losses}
A third family learns a class representative jointly with the encoder. Center loss adds a pull term $\tfrac{1}{2}\sum_i\|f_w(x_i)-\mathbf{c}_{y_i}\|^2$ toward a learned class center \citep{wen2016discriminative}, and ArcFace normalizes embeddings and class weights onto the hypersphere and adds a margin $m$ to the angle $\vartheta_{y_i}$ between an embedding and its class direction \citep{deng2019arcface},
\begin{equation} \label{eq:arcface}
    \mathcal{L}_{\mathrm{ArcFace}} = -\log\frac{\exp\bigl(\kappa\cos(\vartheta_{y_i}+m)\bigr)}{\exp\bigl(\kappa\cos(\vartheta_{y_i}+m)\bigr)+\sum_{k\neq y_i}\exp\bigl(\kappa\cos\vartheta_k\bigr)} ,
\end{equation}
with scale $\kappa$. Both require a class representative and an angle that behaves additively, neither of which is automatic when the target space is a product of projective spaces and the similarity a product of per-factor overlaps.

\subsection{Quantum States, Gates and Encodings} \label{appendix:qml_background}

\paragraph{State space.} An $N$-qubit register has state space $(\C^2)^{\otimes N}$. States are unit vectors defined up to a global phase, so pure states live on $\mathbb{CP}^{2^N-1}$, a compact manifold of real dimension $2(2^N-1)$ carrying the fidelity and Fubini--Study distance of Eq.~\ref{eq:fidelity}. Fidelity is invariant under any unitary applied to both arguments, $F(V\ket{\psi},V\ket{\phi}) = F(\ket{\psi},\ket{\phi})$, which is the identity behind Eq.~\ref{eq:terminal_isometry}.

\paragraph{The single-qubit and product manifolds.} For $N=1$, $\mathbb{CP}^1$ is identified with the Bloch sphere $S^2$: a state is a real unit vector $\mathbf{r}\in\R^3$, its fidelity with another is $F = \tfrac{1}{2}(1+\mathbf{r}\cdot\mathbf{r}')$, and $d_{\mathrm{FS}}$ is half the angle between the two Bloch vectors, so $\mathbb{CP}^1$ is isometric to a sphere of radius $\tfrac{1}{2}$, so a single qubit behaves like a normalized classical embedding. States that factorize as $\ket{\psi} = \bigotimes_{q=1}^{N}\ket{\psi_q}$ carry no entanglement and occupy the product manifold $\MU \cong (\mathbb{CP}^1)^N$, of real dimension $2N$ rather than $2(2^N-1)$. On $\MU$ the fidelity is a product of single-qubit overlaps (Lemma~\ref{lem:factorized_inner_product}), computable classically in $\mathcal{O}(N)$.

\begin{definition}[Fubini--Study ball] \label{def:fs_ball}
Let $\M$ be $\mathbb{CP}^{2^N-1}$ or $\MU$, and let $\mu$ be its uniform probability measure: the unitarily invariant (Haar) measure on $\mathbb{CP}^{2^N-1}$, and its $N$-fold product over qubits on $\MU$. The Fubini--Study ball of radius $t \in [0,\tfrac{\pi}{2}]$ about $\psi \in \M$ is
\begin{equation*}
    \mathcal{B}_t(\psi) \;=\; \bigl\{\phi \in \M : d_{\mathrm{FS}}(\psi,\phi) \leq t\bigr\} \;=\; \bigl\{\phi \in \M : F(\psi,\phi) \geq \cos^2 t\bigr\} ,
\end{equation*}
and the fraction of $\M$ it occupies is $\mu\bigl(\mathcal{B}_t(\psi)\bigr)$, the probability that a uniformly random state lies within distance $t$ of $\psi$. Invariance of $\mu$ makes this fraction independent of $\psi$, so we write $\mu(\mathcal{B}_t)$.
\end{definition}

\paragraph{Gates as rotations.} Up to global phase, every single-qubit unitary is a rotation of the Bloch sphere: $\exp(-i\vartheta P/2)$ for a Pauli operator $P$ rotates the Bloch vector by $\vartheta$ about the $P$ axis. The two rotations used in this work are
\begin{equation} \label{eq:ry_rz}
    R_Y(\theta) = \begin{pmatrix} \cos\frac{\theta}{2} & -\sin\frac{\theta}{2} \\ \sin\frac{\theta}{2} & \cos\frac{\theta}{2} \end{pmatrix}, \qquad
    R_Z(\gamma) = \begin{pmatrix} e^{-i\gamma/2} & 0 \\ 0 & e^{i\gamma/2} \end{pmatrix},
\end{equation}
and applied in sequence to $\ket{0}$ they prepare
\begin{equation} \label{eq:ry_rz_state}
    R_Z(\gamma)R_Y(2\theta)\ket{0} = e^{-i\gamma/2}\bigl(\cos\theta\,\ket{0} + e^{i\gamma}\sin\theta\,\ket{1}\bigr),
\end{equation}
which, dropping the global phase, is the polar form of a qubit with polar angle $2\theta$ and azimuth $\gamma$. Two rotations about distinct axes therefore reach all of $\mathbb{CP}^1$ with two real parameters, and applied qubit-wise they prepare the product state of Eq.~\ref{eq:qpmel_state}. Local rotations never leave $\MU$; reaching the rest of $\mathbb{CP}^{2^N-1}$ requires entangling gates such as CNOT, CZ or $ZZ(\alpha)$.

\paragraph{Parameterized quantum circuits.} A parameterized quantum circuit (PQC) alternates layers of single-qubit rotations with fixed entangling blocks,
\begin{equation} \label{eq:pqc}
    U(\vartheta) = \prod_{l=1}^{L} W_l\,V_l(\vartheta_l), \qquad V_l(\vartheta_l) = \bigotimes_{q=1}^{N}\exp\bigl(-i\,\vartheta_{l,q}P_{l,q}/2\bigr),
\end{equation}
and prepares $U(\vartheta)\ket{0}^{\otimes N}$. In variational models $\vartheta$ is trained by gradient descent, with hardware gradients from the parameter-shift rule \citep{MitaraiParamShift2018,mcclean2016theory,Cerezo_2021}. The layer count, the entangling pattern in $W_l$ and the generators $P_{l,q}$ together fix the reachable states, and with them the expressivity and trainability of Section~\ref{appendix:trainability_expressivity}.

\paragraph{Encoding classical data.} Angle encoding writes each real feature into the angle of one rotation, $\ket{x} = R_Y(x)\ket{0} = \cos(x/2)\ket{0} + \sin(x/2)\ket{1}$. A single rotation axis reaches only a great circle of each Bloch sphere, which is the access the encoders of Section~\ref{subsubsec:enc_vqc} give the classical network. A second rotation about a different axis, as in Eq.~\ref{eq:ry_rz_state}, gives the \emph{dense} angle encoding of \citet{LaRose2020}, with two real features per qubit and access to the full sphere; this is the parameterization \qpmel{} uses. Interleaving encoding and variational layers instead re-uploads the data \citep{DataReupload}, the family of Section~\ref{subsubsec:reupload}.

\paragraph{Estimating fidelity on hardware.} Fidelity is not an observable and must be estimated from repeated measurements. The \emph{inversion} (compute--uncompute) test prepares $\ket{\psi(x)}$ on $N$ qubits, applies the inverse of the circuit preparing $\ket{\psi(x')}$, and estimates the probability of the all-zero outcome, which equals $F$. The \emph{SWAP} test prepares the two states on separate registers, applies controlled-SWAPs from an ancilla, $2N+1$ qubits in all, and recovers $F = 2P(0)-1$ from the ancilla alone. Both estimate a Bernoulli parameter from $S$ shots, so the standard error of $\hat{F}$ is $\mathcal{O}(S^{-1/2})$ independent of $N$, while the fidelities to be resolved shrink with $N$ (Section~\ref{appendix:trainability_expressivity}). Appendix~\ref{app:fsnr_derivations} quantifies the resulting shot budget.

\subsection{Quantum Kernel Models} \label{appendix:quantum_kernels}
A feature map $x \mapsto \ket{\psi(x)}$ defines the kernel $k(x,x') = |\braket{\psi(x)|\psi(x')}|^2$, which is positive semi-definite by construction: the overlaps form the Gram matrix of unit vectors, and taking the squared modulus entrywise is the Schur product of that matrix with its conjugate, which preserves positive semi-definiteness \citep{Schuld_2019,schuld2021supervised}. Quantum kernel methods estimate this Gram matrix on a device and fit a classical convex model, such as an SVM, on top of it \citep{Havl_ek_2019}. The feature map need not be fixed: quantum embedding kernels tune the encoding by maximizing kernel-target alignment \citep{cristianini2001kernel,Cortes_2012,Hubregtsen_2022}, a metric-learning objective stated on the Gram matrix, and \citet{lloyd2020quantum} train an embedding to separate classes by fidelity.

\paragraph{Variational models are kernel methods.} Expanding an observable in its eigenbasis, $O = \sum_k\lambda_k\ket{v_k}\bra{v_k}$, the measured value is a fixed combination of fidelities against reference states,
\begin{equation} \label{eq:observable_as_kernel}
    \braket{\psi(x)|O|\psi(x)} = \sum_k\lambda_k\,\bigl|\braket{v_k|\psi(x)}\bigr|^2 = \sum_k\lambda_k\,k(v_k,x),
\end{equation}
and a variational block $V$ before the measurement only replaces $O$ by $V^\dagger O V$, rotating the $\ket{v_k}$ without leaving this form. The encoding, not the variational block, therefore fixes the hypothesis class \citep{schuld2021supervised}, although the two families can differ in what they learn efficiently in practice \citep{Jerbi_2023}. The parameter-predicting setup of Eq.~\ref{eq:analytical_setup} is a kernel of this kind whose feature map is learned by a classical network, the quantum analogue of a deep embedding kernel \citep{le2018deep}.

\subsection{Trainability and Expressivity} \label{appendix:trainability_expressivity}

\paragraph{Expressivity and the dynamical Lie algebra.} The expressivity of an ansatz is usually measured by how uniformly its outputs cover the state space: an ensemble matching the first two Haar moments is an approximate $2$-design \citep{Harrow_2009}, a regime deep hardware-efficient circuits reach quickly \citep{holmes2022connecting}, and capacity measures such as the effective dimension tell the same story \citep{abbas2021power}. The dynamical Lie algebra (DLA) measures the same quantity from the ansatz alone \citep{larocca2022diagnosing,Ragone_2024}. Write the gates of Eq.~\ref{eq:pqc} as $\exp(-i\vartheta_l H_l)$ with Hermitian generators $H_l$. Two such gates compose into a third of the same form only when their generators commute; otherwise the Baker--Campbell--Hausdorff expansion
\begin{equation} \label{eq:bch}
    e^{A}e^{B} = \exp\Bigl(A + B + \tfrac{1}{2}[A,B] + \tfrac{1}{12}\bigl[A,[A,B]\bigr] - \tfrac{1}{12}\bigl[B,[A,B]\bigr] + \dots\Bigr)
\end{equation}
shows that repeated layers open directions given by nested commutators of the generators. The DLA $\mathfrak{g} \subseteq \mathfrak{su}(2^N)$ of Eq.~\ref{eq:dla} is the Lie closure of $\{iH_l\}$ and is computed by a fixed-point iteration: start from $B_0 = \{iH_1,\dots,iH_M\}$, form all commutators of pairs of current elements, expand each in the Pauli basis, keep those linearly independent of the span accumulated so far, and repeat until a pass adds nothing. The iteration terminates because $\dim\mathfrak{su}(2^N) = 4^N-1$ bounds the span, and $\dim\mathfrak{g}$ is the size of the resulting basis.

The reachable unitaries form the connected group $e^{\mathfrak{g}}$, so the reachable states are the orbit of $\ket{0}^{\otimes N}$ under it, a manifold of dimension at most $\dim\mathfrak{g}$. Three cases bracket the range. Mutually commuting generators give an abelian $\mathfrak{g}$ of dimension at most $M$ whose orbit is a torus. Single-qubit rotations on every qubit give $\mathfrak{g} = \bigoplus_q\mathfrak{su}(2)$ of dimension $3N$, whose orbit is exactly $\MU$, each factor being the Bloch sphere. Generic entangling layers close the algebra to $\mathfrak{su}(2^N)$, whose orbit is all of $\mathbb{CP}^{2^N-1}$. Structured ansätze sit in between; free-fermion and transverse-field Ising generators close into polynomially sized algebras and are the standard way to keep $\dim\mathfrak{g}$ tractable \citep{larocca2022diagnosing}.

\paragraph{Barren plateaus.} For a cost $C(\vartheta) = \braket{\psi(\vartheta)|O|\psi(\vartheta)}$, a barren plateau means $\E_{\vartheta}[\partial_\mu C] = 0$ and $\Var_{\vartheta}[\partial_\mu C] \in \mathcal{O}(b^{-N})$ for some $b > 1$ \citep{mcclean2018barren}. Chebyshev's inequality,
\begin{equation} \label{eq:bp_chebyshev}
    \Pr\bigl[\,|\partial_\mu C| \geq \delta\,\bigr] \leq \frac{\Var_{\vartheta}[\partial_\mu C]}{\delta^2},
\end{equation}
then confines the gradient to within $\mathcal{O}(b^{-N/2})$ of zero with high probability, and resolving a gradient of that size from finite shots costs a budget growing as $\mathcal{O}(b^{N})$. Vanishing variance alone is not a plateau: a gradient concentrated tightly around a nonzero value is still trainable, so a plateau also requires the concentration point to be zero and the cost itself to concentrate. When the generators, observable, and input state lie in the dynamical Lie algebra $\mathfrak{g}$, the variance scales as $\Var_{\vartheta}[\partial_\mu C] \propto \mathcal{P}_{\mathfrak{g}}(\rho)\,\mathcal{P}_{\mathfrak{g}}(O)/\dim\mathfrak{g}$ for input state $\rho$, where the $\mathfrak{g}$-purity $\mathcal{P}_{\mathfrak{g}}$ is the squared norm of its projection onto $\mathfrak{g}$ \citep{Ragone_2024}. On the full algebra $\mathfrak{su}(2^N)$, a global rank-one projector observable has purity $\Theta(1)$, yielding $\Var \in \Theta(4^{-N})$ matching Haar integration \citep{cerezo2021cost}. On restricted subalgebras such as $\bigoplus_q \mathfrak{su}(2)$, however, the global projector has components up to weight $N$ outside $\mathfrak{g}$, yielding purities that vanish exponentially ($\mathcal{P}_\mathfrak{g} = N 2^{-N}$); as proved by \citet{cerezo2021cost}, global fidelity costs suffer exponential barren plateaus ($\Theta(c^{-N})$) across both unentangled and entangled spaces (Lemma~\ref{lem:capacity_variance}). For local observables acting on few-body subsystems ($k \ll N$), by contrast, polynomial trainability is preserved \citep{cerezo2021cost}. In a hybrid model the vanishing gradient sits between the loss and the classical encoder, so it degrades the trainability of the classical parameters as well.

\paragraph{Fidelity concentration.} Exponential concentration is the statement that $\Pr[\,|k(x,x')-\mu| \geq \delta\,] \leq \epsilon$ with $\delta \in \mathcal{O}(b^{-N})$ for essentially all pairs of inputs \citep{thanasilp2024exponential}. It is indexed by the data rather than the parameters, so it persists where gradients are well behaved and survives when the circuit has no trainable parameters at all; expressive embeddings, entanglement, hardware noise and qubit count each drive it. For Haar-random states on $\mathbb{CP}^{2^N-1}$, $F$ is $\mathrm{Beta}(1,2^N-1)$ distributed (Lemma~\ref{lem:ball_volume}), so
\begin{equation} \label{eq:haar_moments}
    \E[F] = 2^{-N}, \qquad \E[F^2] = \frac{2}{2^N(2^N+1)}, \qquad \Var[F] = \Theta(4^{-N}).
\end{equation}
The Gram matrix $K_{ij} = k(x_i,x_j)$ then approaches the identity: its off-diagonal entries become nearly equal, so a loss cannot tell which pairs are more similar. A model can still fit its training set, because the diagonal stays at $1$, but only by memorizing it.

The product manifold is not exempt. For two independent uniform points on a Bloch sphere $\mathbf{r}\cdot\mathbf{r}'$ is uniform on $[-1,1]$, so $F_q = \tfrac{1}{2}(1+\mathbf{r}\cdot\mathbf{r}')$ is uniform on $[0,1]$, and the product $F = \prod_q F_q$ of $N$ independent factors has $\E[F] = 2^{-N}$ and $\E[F^2] = 3^{-N}$, hence
\begin{equation} \label{eq:product_concentration}
    \E[F] = 2^{-N}, \qquad \Var[F] = 3^{-N} - 4^{-N} = \Theta(3^{-N}).
\end{equation}
Uniformly spread product states therefore concentrate around the same value as Haar-random states, with a variance larger by a factor approaching $(4/3)^N$: removing entanglement changes how tightly the fidelities gather, not where they gather. On hardware the estimator noise does not shrink with $N$ while the separations to be resolved do, so distinguishing two kernel entries costs shots growing exponentially in $N$ \citep{miroszewski2024shots,Wang_2021}.

%% file: Appendix/Analysis_extended.tex
This appendix collects the arguments that Section~\ref{section:analysis} states in brief.

\subsection{The setup bounds the families it replaces} \label{app:upper_bound}
Every pure-state model considered here maps a fixed input to $\ket{\psi(x)} = U(z)\ket{0}^{\otimes N}$, so the only thing an ansatz contributes is its reachable set $\mathcal{R}(U) = \{U(z)\ket{0}^{\otimes N} : z \in \Theta\}$. When $f_w$ supplies $z$ per input, any embedding whose image lies in $\mathcal{R}(U)$ is representable up to the approximation error of the network, so two ansätze with the same reachable set are interchangeable here. The families of Sections~\ref{subsubsec:enc_vqc} and~\ref{subsubsec:reupload} hold their trainable parameters fixed across inputs, so each realizes one particular map into the same set, and $f_w$ can output those parameter values itself; the converse fails, since their parameters cannot vary with the input. Results obtained in this setup therefore bound what either family can achieve on the same ansatz. The equivalence concerns representable embeddings only: the parameterization still determines the gradients that reach $f_w$.

\subsection{The packing argument} \label{app:packing_argument}
Place a Fubini--Study ball $\mathcal{B}_{\varepsilon/2}$ of radius $\varepsilon/2$ (Definition~\ref{def:fs_ball}) around each state of an $\varepsilon$-separated set. The balls are disjoint, since overlapping balls would place their centres closer than $\varepsilon$, so the number of states is at most $1/\mu(\mathcal{B}_{\varepsilon/2})$, where $\mu$ is the fraction of the volume of $\M$ that a ball occupies. If the set is as large as possible, no further state can be added, so every point of $\M$ lies within $\varepsilon$ of some state in the set; the balls of radius $\varepsilon$ then cover $\M$, and the number of states is at least $1/\mu(\mathcal{B}_{\varepsilon})$. On $\mathbb{CP}^{2^N-1}$ the ball fraction is known exactly (Appendix~\ref{app:capacity}),
\begin{equation} \label{eq:cap_area}
    \mu(\mathcal{B}_t) = (\sin t)^{m} , \qquad m = \dim_{\R}\M ,
\end{equation}
where $m$ is the number of real parameters of a state: $m = 2(2^N-1)$ on $\mathbb{CP}^{2^N-1}$, and $m = 2N$ on $\MU$, a polar and an azimuthal angle per qubit. Eq.~\ref{eq:cap_area} is the Euclidean law that the volume of an $m$-dimensional ball grows as the $m$-th power of its radius, with the arc length $t$ replaced by $\sin t = \sqrt{1-F}$, the length of the component of one state orthogonal to the other. For small $t$, $\sin t \approx t$; at $t = \pi/2$ the ball is the whole space and the count degenerates. On the product manifold the same law holds only up to a correction of order $N\log N$, which Lemma~\ref{lem:packing} carries as $\Delta$.

Both bounds follow by substituting Eq.~\ref{eq:cap_area} into the two counts above, and both have the same form: the log-capacity of a manifold is its dimension times a constant set by the resolution. At $\varepsilon = \pi/2$, Eq.~\ref{eq:packing_bounds} degenerates and the capacity is the orthogonal count $2^N$ of Eq.~\ref{eq:orthogonal_capacity}. At $\varepsilon = \pi/4$, where two states still have fidelity $\tfrac{1}{2}$, $1/\sin\varepsilon = \sqrt{2}$ and the lower bound becomes $m\log_2\sqrt{2} = m/2$, so $M(\M,\pi/4) \geq 2^{m/2}$: each of the $m/2$ complex dimensions doubles the number of states the bound guarantees. The full space has $2^N-1$ complex dimensions, so $M(\mathbb{CP}^{2^N-1},\pi/4) \geq 2^{2^N-1}$, with a matching upper bound of $2^{2.77(2^N-1)}$. Halving the separation from $\pi/2$ to $\pi/4$ thus raises the capacity of the full space from $2^N$ to exponential in $2^N$.

The product manifold grows far more slowly. A single qubit has six states pairwise at least $\pi/4$ apart, the octahedron $\{\ket{0},\ket{1},\ket{\pm},\ket{\pm i}\}$, so $M(\MU,\pi/4) \geq 6^N$, while Eq.~\ref{eq:packing_bounds} caps it at $(4+2\sqrt{2})^{N} N^{N}$. At $N = 10$ the product manifold holds at most $10^{19}$ such states, where the full space holds at least $10^{307}$, and each further halving of $\varepsilon$ multiplies both counts by roughly $2^{m}$. The proof is given in Appendix~\ref{app:capacity}.

\subsection{Closure of the dynamical Lie algebra} \label{app:closure}
Reaching more states requires a larger DLA $\mathfrak{g}$ (Section~\ref{subsubsec:capacity}). When the observable or input state lies in the circuit's algebra, gradient variance scales inversely with the DLA dimension as $1/\dim\mathfrak{g}$ \citep{Ragone_2024}; this condition fails for global fidelity on $\MU$ (Appendix~\ref{app:capacity}). For global fidelity observables, however, gradients collapse exponentially across both unentangled and entangled spaces (Lemma~\ref{lem:capacity_variance}), and reaching all of $\mathbb{CP}^{2^N-1}$ with $\mathfrak{g} = \mathfrak{su}(2^N)$ incurs $\Theta(4^{-N})$ variance \citep{cerezo2021cost}.

For $\mathfrak{su}(2^N)$, Eq.~\ref{eq:dla_quadratic} is quadratic in the dimension reached rather than proportional to it. The two terms split the algebra at a given state: $m$ directions move the state and the remaining $(m/2)^2$ leave it unchanged (they form its stabilizer), so all but $m$ of the $4^N-1$ directions are invisible to the embedding.

Generic entanglement reaches $\mathfrak{su}(2^N)$ because an algebra is closed under commutators. In the Pauli basis, $\mathfrak{su}(2^N)$ has one generator per non-identity Pauli string, $4^N-1$ of them, and local rotations occupy only the $3N$ strings of weight one (those acting on a single qubit). The commutator of two anticommuting Pauli strings is a third, so two-qubit couplings across a connected set of pairs generate strings of weight two, these generate weight three together with the local rotations, and the closure continues until every string is included. At $N = 10$ local rotations provide $30$ of the $1\,048\,575$ generators, and adding a generic coupling generates all of the rest.

Structured generator sets are the exception: they close on subalgebras of polynomial dimension \citep{larocca2022diagnosing}, and a model that reaches beyond $\MU$ at polynomial gradient cost must use one. These subalgebras form a discrete family rather than a continuum, and each fixes its own reachable submanifold, committing the embedding in advance to the states that submanifold contains. Elsewhere in QML, such a restriction is justified when a good solution is known to lie in the retained subspace; in metric learning, the size of the reachable set is the reason for using the quantum state space, so that justification is unavailable. $\MU$ avoids this at maximal separation, since it retains all $2^N$ orthogonal states (Eq.~\ref{eq:orthogonal_capacity}).

\subsection{Pairwise cost and locality of the product kernel} \label{app:locality}
Classically, $n$ inputs are embedded once and all $n^2$ similarities follow from inner products of the stored vectors. A generic quantum state has no efficient classical description, so each pair needs its own circuit execution: $\mathcal{O}(n^2)$ circuits per batch at $S$ shots each. A CLIP-style batch of $10^4$ contains $5 \times 10^{7}$ pairs, or $5 \times 10^{10}$ circuit executions per step at $10^3$ shots. Retrieval is worse: scoring a query against a database costs $\mathcal{O}(|\mathrm{DB}|)$ circuit runs with no index to prune them, whereas classical search is sublinear \citep{malkov2018efficient}. Both costs revert to classical ones when the state is classically tractable, since on $\MU$ Lemma~\ref{lem:factorized_inner_product} returns $F$ in $\mathcal{O}(N)$ arithmetic.

For small misalignments the factorized fidelity is a Gaussian kernel. Writing $\beta_q$ for the Bloch angle between the two states of qubit $q$,
\begin{equation} \label{eq:rbf_limit}
    F = \prod_{q} \cos^{2}\!\frac{\beta_q}{2} \;\approx\; \exp\Bigl(-\tfrac{1}{4}\textstyle\sum_{q} \beta_q^{2}\Bigr) ,
\end{equation}
accurate to under a percent for misalignments near $0.1$ radians. Fidelity is thus a Gaussian (RBF) kernel whose bandwidth is set by the output scale of $f_w$ rather than chosen. Local kernel machines need a number of samples exponential in the input dimension to represent functions that vary on the kernel's own scale \citep{bengio2005curse}, and the concentration of Section~\ref{subsubsec:dnn_kernel} drives the Gram matrix toward the identity, shrinking the effective bandwidth further. The effective degrees of freedom then equal the number of training points, and no non-trivial generalization bound survives. Shattering is therefore the standard failure of a kernel with too small a bandwidth: the exponentially many orthogonal states of the space make such a bandwidth reachable, and an expressive encoder reaches it. Removing those states would also remove the capacity classification needs (Appendix~\ref{subsubsec:smoothness}), so the remedy lies in the encoder.

\subsection{Trivial minimizers and the choice of loss} \label{subsubsec:minimizers}
A \emph{trivial minimizer} is an arrangement of embeddings that attains the objective's minimum while encoding nothing about the data. Metric-learning losses are computed from the similarities inside tuples of embeddings, pairs for contrastive losses, triplets for the hinge, the whole batch for InfoNCE (Appendix~\ref{appendix:metric_learning}), so each reads a batch only through its Gram matrix $K_{ij} = k_w(x_i,x_j)$, with proxy losses adding the similarities to a set of class representatives. Setting $K = I$ places every negative pair at maximal distance and so minimizes any objective built from negatives whatever the samples were, and every orthonormal $n$-frame inside $\MU$ realizes it (Proposition~\ref{prop:shattering}), so such arrangements are abundant rather than rare.

\paragraph{Saturating and non-saturating losses reach $K = I$ differently.} The triplet hinge contributes gradient only while a negative violates the margin, and under concentration the typical negative starts at $d_{\mathrm{FS}} \approx \pi/2$, leaving the term inactive before training begins. InfoNCE has no margin and its gradient vanishes only with the negative similarities themselves, so it pushes toward orthogonality for as long as any remains. This predicts which tasks shatter: class-level objectives on MNIST and Omniglot do not, whereas under InfoNCE a sufficiently expressive encoder does on THINGS-EEG2 (Section~\ref{subsec:trainability_results}).

\paragraph{A class representative is not a mean.} The representatives that prototypical, center and ArcFace losses require must themselves be states, and an arithmetic mean of state vectors is not one: it is neither normalized nor phase-invariant, since $\ket{\psi}$ and $e^{i\alpha}\ket{\psi}$ are the same state and average differently. The fidelity-optimal prototype maximizes $\tfrac{1}{n}\sum_i |\braket{\phi|\psi_i}|^2 = \bra{\phi}\bar{\rho}\ket{\phi}$ for $\bar{\rho} = \tfrac{1}{n}\sum_i \ket{\psi_i}\!\bra{\psi_i}$, and is its top eigenvector; restricted to $\MU$ the same maximization is a best rank-one tensor approximation, with no closed form and NP-hard in general. The additive surrogate $\sum_q F_q$ separates over qubits and is maximized by the top eigenvector of each reduced state $\bar{\rho}_q$, the per-qubit Bloch mean, a second reason to prefer it to the product kernel (Section~\ref{subsec:curriculum}).

\subsection{Smoothness} \label{subsubsec:smoothness}
Smoothness requires the composite map $x \mapsto \ket{\psi_w(x)}$ to respond proportionally to its input. A small change in the encoder's output must move $d_{\mathrm{FS}}$ by a small amount, so that gradients remain informative, and inputs close in $\mathcal{X}$ must stay close in $\M$ while distinct inputs stay distinct, so that a margin won in training transfers to unseen points. The geometry enforces neither condition.

\paragraph{The parameterization is periodic.} Gate angles enter as $\exp(-izP/2)$, and the state is unchanged (up to a global phase) by $z \mapsto z + 2\pi$, so $z \mapsto U(z)\ket{0}^{\otimes N}$ is only locally injective. An encoder with unbounded outputs therefore aliases: inputs arbitrarily far apart in $\mathcal{X}$ can be assigned the same state, and the objective reads that pair at fidelity one, as a positive.

\paragraph{The Lipschitz constant is unconstrained.} Let $\kappa$ be the smallest constant with $d_{\mathrm{FS}}(\psi_w(x),\psi_w(x')) \leq \kappa\, d_{\mathcal{X}}(x,x')$ for all $x,x'$. The geometry bounds the left side by $\pi/2$ but places no bound on $\kappa$, since $f_w$ may cross that diameter over an arbitrarily small input distance and send neighbouring inputs to orthogonal states. A margin won in training therefore certifies nothing between the training points: the embedding can satisfy every constraint on $\mathcal{D}$ while varying arbitrarily fast elsewhere, which is geometric shattering (Proposition~\ref{prop:shattering}) stated as a property of the map. Generalization bounds for quantum embeddings scale with $\kappa$ rather than with the dimension of the Hilbert space \citep{kim2022generalization}, so $\kappa$ is the quantity a design must control. Bounding it also removes the aliasing above, by capping the angle a bounded input range can sweep.

\paragraph{Shattering needs both the geometry and an expressive encoder.} It needs $n$ mutually orthogonal reachable states, which the exponential dimension supplies, and removing entanglement does not remove them: the $2^N$ basis states of Eq.~\ref{eq:orthogonal_capacity} lie inside $\MU$ and local rotations reach every one. It also needs an encoder expressive enough to assign inputs to them. Under the same \qpmel{} kernel and InfoNCE objective, the convolutional QSTConv does not shatter, with $0.2\%$ of unmatched pairs below $0.01$, whereas the unregularized transformer QNervFormer sends a third of them there (Table~\ref{tab:kernel_geometry}). Either factor can be restricted. A severe enough restriction of the circuit removes the orthogonal states, since a permutation-symmetric circuit never leaves the symmetric subspace, which holds only $N+1$ mutually orthogonal states. But this caps the orthogonal capacity at $N+1$, an advance commitment of the kind discussed in Appendix~\ref{app:closure}, rather than a remedy. For any ansatz whose orthogonal capacity exceeds $n$, the constraint must therefore fall on $f_w$, through its architecture or its regularization.

%% file: Appendix/proof.tex
Results are proved in the order the main text uses them: the capacity bounds of Section~\ref{subsubsec:capacity} (Appendix~\ref{app:capacity}), geometric shattering (Appendix~\ref{app:shattering}), the shot-noise scaling of Section~\ref{subsubsec:fidelity} (Appendix~\ref{app:fsnr_derivations}), and the closed-form product-state fidelity used in Sections~\ref{subsubsec:fidelity} and~\ref{section:qpmel} (Appendices~\ref{appendix/eq:InnerProductFactorProof}--\ref{appendix/eq:PMeFProof}).

\subsection{Capacity of the Quantum State Space} \label{app:capacity}

This section proves Lemma~\ref{lem:packing} and Lemma~\ref{lem:capacity_variance}. Throughout, $d = 2^N$, distances are Fubini--Study, and $\mu$ is the unitarily invariant probability measure on the state space, so $\mu(A)$ is the probability that a Haar-random state lies in $A$ and $\mu(\mathcal{B}_t)$ is the fraction of the space covered by a metric ball of radius $t$. Invariance is what the argument uses: every ball of a given radius carries the same measure, wherever its centre.

\begin{lemma}[Fubini--Study ball measure] \label{lem:ball_volume}
For any fixed $\ket{\psi} \in \mathbb{CP}^{d-1}$ and $\ket{\phi}$ drawn from the unitarily invariant measure, $\Pr[F(\psi,\phi) \geq u] = (1-u)^{d-1}$ for $u \in [0,1]$. The measure of the ball $\mathcal{B}_t(\psi)$ of Definition~\ref{def:fs_ball} is therefore
\begin{equation} \label{eq:ball_volume}
    \mu\bigl(\mathcal{B}_t(\psi)\bigr) = \sin^{2(d-1)}(t), \qquad t \in [0,\tfrac{\pi}{2}].
\end{equation}
\end{lemma}
\begin{proof}
By unitary invariance take $\ket{\psi} = \ket{0}$, so $F = |\braket{0|\phi}|^2$ is the first squared component of a uniform point on the unit sphere of $\C^{d}$, which is $\mathrm{Beta}(1,d-1)$ distributed. Its survival function is $\Pr[F \geq u] = (1-u)^{d-1}$. A state lies in $\mathcal{B}_t(\psi)$ exactly when $\arccos\sqrt{F} \leq t$, i.e.\ when $F \geq \cos^2 t$, and substituting $u = \cos^2 t$ gives $\mu(\mathcal{B}_t) = (1-\cos^2 t)^{d-1} = \sin^{2(d-1)}(t)$.
\end{proof}

\begin{proof}[Proof of Lemma~\ref{lem:packing}]
Let $\mathcal{S}$ be a maximal $\varepsilon$-separated subset of $\mathbb{CP}^{d-1}$, so $|\mathcal{S}| = M(\mathbb{CP}^{d-1},\varepsilon)$. The balls $\{\mathcal{B}_{\varepsilon/2}(\psi)\}_{\psi \in \mathcal{S}}$ are pairwise disjoint, since two of them meeting would place their centres within $\varepsilon$ of each other, and they all sit inside a space of unit measure, so $|\mathcal{S}| \, \mu(\mathcal{B}_{\varepsilon/2}) \leq 1$. Maximality means no point can be added, so every point of the space lies within $\varepsilon$ of some centre and $\{\mathcal{B}_{\varepsilon}(\psi)\}_{\psi \in \mathcal{S}}$ covers it, giving $|\mathcal{S}| \, \mu(\mathcal{B}_{\varepsilon}) \geq 1$. Substituting Eq.~\ref{eq:ball_volume} with $d-1 = 2^N-1$ and taking logarithms yields Eq.~\ref{eq:packing_bounds} with $m = 2(2^N-1)$ and $\Delta = 0$.

For the product manifold, $\MU \cong (\mathbb{CP}^1)^N$ carries the product metric and the product measure, and a ball of radius $\varepsilon$ about $\psi = \bigotimes_q \psi_q$ contains the product of the per-qubit balls $\mathcal{B}_{\varepsilon/\sqrt{N}}(\psi_q)$. Eq.~\ref{eq:ball_volume} at $d = 2$ gives $\mu(\mathcal{B}_t) = \sin^2 t$ on each factor, so
\begin{equation*}
    \sin^{2N}\!\left(\frac{\varepsilon}{\sqrt{N}}\right) \;\leq\; \mu\bigl(\mathcal{B}_\varepsilon(\psi)\bigr) \;\leq\; \sin^{2N}(\varepsilon),
\end{equation*}
the upper bound because the ball is contained in the product of the per-qubit balls of radius $\varepsilon$. The same two counting arguments then bracket $\log M(\MU,\varepsilon)$ between $2N \log(1/\sin\varepsilon)$ and $2N \log(\sqrt{N}/\sin(\varepsilon/2))$, which is Eq.~\ref{eq:packing_bounds} with $m = 2N$ and $\Delta = N\log N$, the slack coming from the $\sqrt{N}$ contraction of the per-qubit radius.
\end{proof}

\begin{proof}[Proof of Lemma~\ref{lem:capacity_variance}]
\textbf{Global fidelity barren plateau.} The global fidelity cost is $C_G(\theta) = 1 - F(\psi(\theta),\phi) = 1 - |\braket{\psi(\theta)|\phi}|^2$, which evaluates the global rank-$1$ projector $O_G = \ket{\phi}\bra{\phi}$.

On $\MU$, the state factorizes into independent single-qubit states $\ket{\psi(\theta)} = \bigotimes_{q=1}^N \ket{\psi_q(\theta_q)}$, so the fidelity factorizes as $F = \prod_{q=1}^N F_q(\theta_q)$. For any parameter $\theta_1$ of qubit $1$, the partial derivative is
\begin{equation*}
    \partial_{\theta_1} C_G = -\left(\frac{\partial F_1}{\partial \theta_1}\right) \prod_{q=2}^{N} F_q(\theta_q) .
\end{equation*}
Under independent parameter distributions with $\E[\partial_{\theta_1} F_1] = 0$, the factors are mutually independent, so $\E[\partial_{\theta_1} C_G] = 0$ and the variance factorizes as
\begin{equation*}
    \Var_{\theta}[\partial_{\theta_1} C_G] = \E\Bigl[\bigl(\partial_{\theta_1} F_1\bigr)^2\Bigr] \prod_{q=2}^{N} \E\bigl[F_q^2\bigr] .
\end{equation*}
If single-qubit states are Haar-distributed on each Bloch sphere $\mathbb{CP}^1$, $F_q \sim \mathrm{Beta}(1,1) = \mathrm{Uniform}(0,1)$, giving $\E[F_q^2] = 1/3$ and $\Var_{\theta}[\partial_{\theta_1} C_G] = \E[(\partial_{\theta_1} F_1)^2] \cdot 3^{-(N-1)}$. For single-qubit rotations $R_y(\theta_q)$ with uniform angles $\theta_q \in [-\pi, \pi]$ against $\ket{0}^{\otimes N}$ (the warm-up example of \citet{cerezo2021cost}), $F_q = \cos^2(\theta_q/2)$, giving $\E[F_q^2] = 3/8$, $\E[(\partial_{\theta_1} F_1)^2] = 1/8$ and $\Var_\theta[\partial_{\theta_1} C_G] = \frac{1}{8}(3/8)^{N-1}$. In the parameterization $R_y(2\theta_q)$ of Eq.~\ref{eq:qpmel_state}, $F_q = \cos^2\theta_q$, and the prefactor becomes $\frac{1}{2}$ with the same base. Thus, on $\MU$, $\Var_\theta[\partial_\mu C_G] = \Theta(c^{-N})$ with $c = 1/\E[F_q^2] > 1$ set by the angle distribution and the target ($c = 3$ for Haar-distributed single-qubit states, $c = 8/3$ above), vanishing exponentially with $N$ despite zero entanglement. The bound concerns uniformly random parameters: a narrow initialization, or a trained encoder whose positives are aligned, has $\E[F_q^2] \to 1$ and escapes it. Moreover, by Proposition~1 of \citet{cerezo2021cost}, for $\delta \in (0,1)$,
\begin{equation*}
    \Pr\bigl[C_G \leq \delta\bigr] = \Pr\left[\prod_{q=1}^N \cos^2\frac{\theta_q}{2} \geq 1-\delta\right] \leq (1-\delta)^{-1}\left(\frac{1}{2}\right)^N ,
\end{equation*}
so the volume of parameter space within distance $\delta$ of the global minimum shrinks exponentially, forming a narrow gorge.

Note that the Lie-algebraic variance formula of \citet{Ragone_2024}, $\Var_\theta[\partial_\mu C] \propto \sum_j \mathcal{P}_{\mathfrak{g}_j}(\rho)\mathcal{P}_{\mathfrak{g}_j}(O)/\dim\mathfrak{g}_j$, requires $\rho \in i\mathfrak{g}$ or $O \in i\mathfrak{g}$. For $\mathfrak{g} = \bigoplus_q \mathfrak{su}(2)$, $i\mathfrak{g}$ contains only $1$-body Pauli operators. The rank-$1$ state projector $\rho = \ket{0}\bra{0}^{\otimes N} = 2^{-N}\sum_S Z_S$ and observable $O_G = \ket{\phi}\bra{\phi}$ have components up to weight $N$, so neither lies in $i\mathfrak{g}$, and their formal purities are $\mathcal{P}_\mathfrak{g}(\rho) = \mathcal{P}_\mathfrak{g}(O_G) = N\cdot 2^{-N}$, which vanish exponentially rather than remaining $\Theta(1)$.

On $\mathbb{CP}^{2^N-1}$, generic entangling circuits close to $\mathfrak{su}(2^N)$ with $\dim\mathfrak{g} = 4^N-1$. When the differentiated gate sits between two independent unitary $2$-designs on $\mathrm{SU}(2^N)$, both the traceless parts of $\rho$ and $O_G$ lie in $i\mathfrak{su}(2^N)$ with purities $\mathcal{P}_\mathfrak{g} = 1 - 2^{-N} = \Theta(1)$. Exact Haar integration over $\mathrm{SU}(2^N)$ yields $\Var_\theta[\partial_\mu C_G] = \frac{1}{2(2^N+1)^2} = \Theta(4^{-N})$ for Pauli generators \citep{mcclean2018barren,cerezo2021cost}.

\textbf{Measurement dimension and local observable trainability.} By Theorem~2 and Corollary~2 of \citet{cerezo2021cost}, when the observable $O$ is an $m$-local operator acting on $k$-qubit subsystems with $k \ll N$, the gradient variance for shallow circuits is lower-bounded by $\Omega(1/\mathrm{poly}(N))$. For the additive kernel $\bar{k}_w(x,x') = \frac{1}{N}\sum_{q=1}^N F_q(x,x')$ introduced in Section~\ref{subsec:curriculum}, the corresponding observable $O_L = \frac{1}{N}\sum_{q=1}^N \ket{\psi_q(x')}\bra{\psi_q(x')} \otimes \mathbb{I}_{\bar{q}}$ is strictly $1$-local ($k = 1$). Differentiating yields $\partial \bar{k}_w/\partial \theta_q = \frac{1}{N} \partial F_q/\partial \theta_q$, so
\begin{equation*}
    \Var_\theta\left[\frac{\partial \bar{k}_w}{\partial \theta_q}\right] = \frac{1}{N^2} \Var_\theta\left[\frac{\partial F_q}{\partial \theta_q}\right] = \Theta\left(\frac{1}{N^2}\right) ,
\end{equation*}
which scales polynomially with $N$ and avoids the barren plateau, confirming the trainability of the local observable regime.
\end{proof}

\paragraph{Tightness.} The volumetric lower bound is not tight on $\MU$. The six octahedral states $\{\ket{0},\ket{1},\ket{\pm},\ket{\pm i}\}$ of a single qubit are pairwise at fidelity at most $\tfrac{1}{2}$, so any two distinct elements of their $N$-fold product differ on some qubit and have $F = \prod_q F_q \leq \tfrac{1}{2}$, giving $M(\MU,\pi/4) \geq 6^N$ where the volumetric bound returns $2^N$. Products of separated sets remain separated under a product fidelity, so constructions of this kind improve the constant but leave the growth of $\log M$ with $m$, which is all Lemma~\ref{lem:packing} is used for, unchanged.

\subsection{Geometric Shattering} \label{app:shattering}

\begin{proof}[Proof of Proposition~\ref{prop:shattering}]
(i) By Lemma~\ref{lem:factorized_inner_product}, $F = \prod_q F_q$ with each $F_q \in [0,1]$. A product of factors in $[0,1]$ is zero if and only if some factor is zero, and one if and only if every factor is one.

(ii) Write $s_{ij} = k_w(x_i,y_j) \in [0,1]$. The $i$-th term of Eq.~\ref{eq:infonce} is
\begin{equation*}
    \ell_i = \log\Bigl(1 + \sum_{j \neq i} \exp\bigl((s_{ij} - s_{ii})/\tau\bigr)\Bigr) ,
\end{equation*}
which depends on the embeddings only through the gaps $s_{ij} - s_{ii}$. Each gap is at least $-1$, so $\ell_i \geq \log\bigl(1 + (B-1)e^{-1/\tau}\bigr)$, with equality if and only if $s_{ii} = 1$ and $s_{ij} = 0$ for every $j \neq i$. Averaging over $i$ gives the bound, attained exactly when $s_{ij} = \delta_{ij}$; the same argument applies to the columns of a symmetric loss. Now $s_{ii} = 1$ means $f_w(x_i)$ and $g_w(y_i)$ are the same state $\psi_i$, and $s_{ij} = 0$ for $i \neq j$ then means $\psi_1,\dots,\psi_B$ are mutually orthogonal. A set of $B$ mutually orthogonal states exists if and only if $B \leq 2^N$ (Eq.~\ref{eq:orthogonal_capacity}), and for $B \leq 2^N$ distinct computational basis states, which lie in $\MU$, realize it. The loss sees only the Gram matrix, which equals the identity for any assignment of samples to orthonormal states, so every such assignment attains the bound. When the training set itself has $n \leq 2^N$ pairs, one assignment attains the bound on every batch simultaneously.

On the unit sphere $S^{d-1}$ with cosine similarity the per-pair floor is $-1$, which holds only for antipodal points. No three points are pairwise antipodal, so for $B > 2$ no batch reaches the floor, and the most even spread of $B$ unit vectors, a regular simplex with pairwise cosine $-1/(B-1)$, needs $d \geq B-1$.
\end{proof}

\subsection{Fidelity Signal-to-Noise Ratio} \label{app:fsnr_derivations}

This section derives the FSNR of Eq.~\ref{eq:fsnr} for the two estimators of Appendix~\ref{appendix:qml_background}. Each estimates $F$ from $S$ shots with a conditional variance $\Var(\hat{F} \mid F)$ fixed by the estimator, and the FSNR follows by substituting the moments of $F$ from Eq.~\ref{eq:haar_moments} for Haar-random states and Eq.~\ref{eq:product_concentration} for uniformly spread product states.

\paragraph{Inversion test.} The all-zero outcome is Bernoulli with parameter $F$, so $\Var(\hat{F} \mid F) = F(1-F)/S$ and
\begin{equation}\label{eqn:fsnr_inv}
    \mathrm{FSNR}_{\mathrm{inv}} = S\,\frac{\E[F^2] - (\E[F])^2}{\E[F] - \E[F^2]}.
\end{equation}
For Haar-random states the numerator is $(d-1)/\bigl(d^2(d+1)\bigr)$ and the denominator $(d-1)/\bigl(d(d+1)\bigr)$, while for uniformly spread product states on $\MU$ the moments are $\E[F] = 2^{-N}$ and $\E[F^2] = 3^{-N}$, so
\begin{equation}\label{eqn:fsnr_haar_inv}
    \mathrm{FSNR}_{\mathrm{inv}}^{\mathrm{Haar}} = S\,2^{-N}, \qquad
    \mathrm{FSNR}_{\mathrm{inv}}^{\mathrm{prod}} = S\,\frac{3^{-N} - 4^{-N}}{2^{-N} - 3^{-N}} = \Theta\bigl(S\,(2/3)^{N}\bigr).
\end{equation}
Holding the ratio constant therefore costs $S \in \Omega(2^N)$ under Haar-random entanglement and $S \in \Omega\bigl((3/2)^N\bigr)$ on the product manifold.

\paragraph{SWAP test.} The ancilla returns $0$ with probability $(1+F)/2$ and $\hat{F} = 2\hat{P}(0) - 1$, so $\Var(\hat{F} \mid F) = (1-F^2)/S$ and
\begin{equation}\label{eqn:fsnr_swap}
    \mathrm{FSNR}_{\mathrm{SWAP}} = S\,\frac{\E[F^2] - (\E[F])^2}{1 - \E[F^2]}.
\end{equation}
For Haar-random states the denominator is $(d-1)(d+2)/\bigl(d(d+1)\bigr)$, giving
\begin{equation}\label{eqn:fsnr_haar_swap}
    \mathrm{FSNR}_{\mathrm{SWAP}}^{\mathrm{Haar}} = \frac{S}{2^N(2^N + 2)} = \Theta\bigl(S\,4^{-N}\bigr),
\end{equation}
so a constant ratio requires $S \in \Omega(4^N)$. These are ideal statistical rates; readout error, gate infidelity and crosstalk on hardware bias $\hat{F}$ and raise the effective cost further.

\subsection{Factorized Inner Product} \label{appendix/eq:InnerProductFactorProof}

\begin{lemma}\label{lem:factorized_inner_product}
For $N$-qubit product states $\ket{\psi_a}=\bigotimes_{i=1}^{N} \ket{\psi_a^i}$ and $\ket{\psi_b}=\bigotimes_{i=1}^{N} \ket{\psi_b^i}$,
\begin{equation*}
    \braket{\psi_a|\psi_b} = \prod_{i=1}^{N} \braket{\psi_a^i|\psi_b^i} .
\end{equation*}
\end{lemma}

\begin{proof}
The adjoint of a tensor product is the tensor product of the adjoints in the same order, $\bigl(\bigotimes_{i}\ket{\psi_a^i}\bigr)^{\dagger} = \bigotimes_{i}\bra{\psi_a^i}$; unlike a matrix product, the factors are not reversed. The mixed-product property $(A_1\otimes\cdots\otimes A_N)(B_1\otimes\cdots\otimes B_N) = A_1B_1\otimes\cdots\otimes A_NB_N$ then gives
\begin{equation*}
    \braket{\psi_a|\psi_b} = \Bigl(\bigotimes_{i=1}^{N} \bra{\psi_a^i}\Bigr)\Bigl(\bigotimes_{i=1}^{N} \ket{\psi_b^i}\Bigr) = \bigotimes_{i=1}^{N} \braket{\psi_a^i|\psi_b^i} = \prod_{i=1}^{N} \braket{\psi_a^i|\psi_b^i},
\end{equation*}
since a tensor product of scalars is their ordinary product.
\end{proof}

\subsection{Complex Kernel Function (CKF)} \label{appendix/eq:ComplexKernelFunctionProof}
\begin{lemma} \label{lem:complex_kernel}
    Let angular encodings $a$ and $b$ prepare the single-qubit states $\ket{\psi_a}$ and $\ket{\psi_b}$, with real coordinate vectors $(x,y,z)$ as in Eq.~\ref{eq:polar_to_cart}. Then
    \begin{equation*}
        |\braket{\psi_a|\psi_b}|^2 = (z_a\cdot z_b + x_a\cdot x_b + y_a \cdot y_b)^2 + (x_a \cdot y_b - y_a \cdot x_b)^2
    \end{equation*}
\end{lemma}

\begin{proof}
    Based on the spherical coordinates we create, we can assume:
    \begin{align*}
        x & = \cos(\gamma)\cdot \sin(\theta) \\
        y & = \sin(\gamma)\cdot \sin(\theta) \\
        z & = \cos(\theta)
    \end{align*}
    As we use only $R_Y$ and $R_Z$ gates, our qubit states can be written as functions of the $x,y,z$ defined above after we expand the exponential term:
    \begin{equation*}
        \ket{\psi} = z\ket{0} + (x+iy) \ket{1}
    \end{equation*}
    Their inner product is therefore
    \begin{align*}
        \braket{\psi_a|\psi_b} & = (z_a\bra{0} + (x_a-iy_a)\bra{1})\cdot (z_b\ket{0} + (x_b + iy_b)\ket{1})             \\
                               & = z_a\cdot z_b + (x_a-iy_a)\cdot(x_b+iy_b)                                             \\
                               & = z_a\cdot z_b + x_a\cdot x_b + i x_a \cdot y_b - i y_a \cdot x_b - i^2  y_a \cdot y_b \\
                               & = (z_a\cdot z_b + x_a\cdot x_b + y_a \cdot y_b) + i(x_a \cdot y_b - y_a \cdot x_b)     \\
                               & = \lambda_r + i \lambda_c
    \end{align*}
    where $\lambda_r = z_a\cdot z_b + x_a\cdot x_b + y_a \cdot y_b$ and $\lambda_c = x_a \cdot y_b - y_a \cdot x_b$ are both real numbers computed using real inner products.
    Applying the magnitude square :
    \begin{align*}
        |\braket{\psi_a|\psi_b}|^2 & =  (\lambda_r + i \lambda_c) \cdot(\lambda_r - i \lambda_c) = \lambda_r^2 - i^2 \lambda_c^2 = \lambda_r^2 + \lambda_c^2 \\
                                   & = (z_a\cdot z_b + x_a\cdot x_b + y_a \cdot y_b)^2 + (x_a \cdot y_b - y_a \cdot x_b)^2
    \end{align*}
    Hence proved.
\end{proof}

\subsection{Projective Metric Function Formula} \label{appendix/eq:PMeFProof}
\begin{lemma} \label{lem:PMeF}
    Given any pair of angular encodings $a,b$ such that they create the $N$-qubit product states $\ket{\psi_a},\ket{\psi_b}$, their similarity in Hilbert Space can be computed via the single qubit CKF (see \ref{appendix/eq:ComplexKernelFunctionProof}) as follows:
    \begin{equation*}
        |\braket{\psi_a|\psi_b}|^2 = \prod_{i=1}^N \left( (z_a^i\cdot z_b^i + x_a^i\cdot x_b^i + y_a^i \cdot y_b^i)^2 + (x_a^i \cdot y_b^i - y_a^i \cdot x_b^i)^2 \right) = \prod_{i=1}^N \left( (\lambda_r^i)^2 + (\lambda_c^i)^2 \right)
    \end{equation*}
\end{lemma}

\begin{proof}
    We know from \ref{appendix/eq:InnerProductFactorProof} -
    \begin{equation*}
        \braket{\psi_a|\psi_b} = \prod_{i=1}^N \braket{\psi_a^i|\psi_b^i} = \braket{\psi_a^1|\psi_b^1} \times \dots \times \braket{\psi_a^N|\psi_b^N}
    \end{equation*}
    Applying the magnitude operator on both sides:
    \begin{equation*}
        |\braket{\psi_a|\psi_b}| = |\braket{\psi_a^1|\psi_b^1} \times \dots \times \braket{\psi_a^N|\psi_b^N}|
    \end{equation*}
    Notice that each inner product $\braket{\psi_a^i|\psi_b^i}$ can be replaced by $\lambda^i \in \C$.
    \begin{equation*}
        |\braket{\psi_a|\psi_b}| = |\lambda^1 \times \dots \times \lambda^N|
    \end{equation*}
    We can then apply the property that $|z_0 \cdot z_1| = |z_0| \cdot |z_1| \;\; \forall z_0,z_1 \in \C$ to get:
    \begin{equation*}
        |\braket{\psi_a|\psi_b}| = |\lambda^1| \times \dots \times |\lambda^N| = |\braket{\psi_a^1|\psi_b^1}| \times \dots \times |\braket{\psi_a^N|\psi_b^N}| = \prod_{i=1}^N |\braket{\psi_a^i|\psi_b^i}|
    \end{equation*}
    Finally, as each magnitude of complex number yields a real number, we can apply the square operator (for real $a_i$, $(a_1\cdots a_N)^2 = a_1^2\cdots a_N^2$) to get:
    \begin{equation*}
        |\braket{\psi_a|\psi_b}|^2 = \prod_{i=1}^N |\braket{\psi_a^i|\psi_b^i}|^2
    \end{equation*}
    Replacing the fidelities with the CKF function calculated earlier in section \ref{appendix/eq:ComplexKernelFunctionProof}:
    \begin{align*}
        |\braket{\psi_a|\psi_b}|^2 & = \prod_{i=1}^N \left( (z_a^i\cdot z_b^i + x_a^i\cdot x_b^i + y_a^i \cdot y_b^i)^2 + (x_a^i \cdot y_b^i - y_a^i \cdot x_b^i)^2 \right) \\
                                   & = \prod_{i=1}^N \left( (\lambda_r^i)^2 + (\lambda_c^i)^2 \right)
    \end{align*}

    Hence proved.
\end{proof}

%% file: Appendix/Results_extended.tex
\subsection{Few-shot baselines} \label{app:fewshot_details}
Baseline figures are those each paper reports; \qpmel{} is evaluated on the same classes, ways and shots, but inputs and encodings are not matched, so the table compares complete designs; the matched-encoder comparison is Table~\ref{tab:eeg}. \citet{thumwanit2021trainable} downsample images to $4 \times 4$ binary pixels and encode them through a trained lookup table of $U_3$ gates, and \citet{wang2024quantum} amplitude-encode $8 \times 8$ images. For \citet{wang2024quantum} Table~\ref{tab:fewshot} reports the strongest model in their table for each task: QDiff, or QuantumNAT on 3-way 1-shot Fashion-MNIST. \qpmel{} uses fewer qubits than the baseline in eight comparisons and the same number in two; the exception is \citet{Huang} on Omniglot (11 against 9). \citet{Liu_2022} feed a CNN's outputs to the metric-learning ansatz of \citet{lloyd2020quantum}, which re-uploads the encoding between trainable $ZZ$ and $R_Y$ layers, and classify by fidelity to class prototypes. Their ablation places an unentangled variant at about $97.8\%$ against $98.3\%$ for the weakly entangled model, and a wider initialization, which entangles more, at about $97.3\%$ (read from their Figures~2 and~3, which report the best test accuracy reached during training).

\subsection{THINGS-EEG2} \label{app:eeg_details}
All heads share the encoders, loss and optimizer of \citet{chen2024quantum}: 10 qubits, AdamW at learning rate $2 \times 10^{-4}$, 200 epochs, batch size 1024 and a learned logit scale initialized at $1/0.07$. Table~\ref{tab:eeg} reports subject 1 over five seeds, except one- and two-layer DR on QNervFormer (ten and three seeds). Its QNervFormer columns are at epoch 200. The QSTConv and branch-dropout columns select the checkpoint with the lowest loss on 740 held-out training pairs, never on the test set; on QSTConv this moves held-out accuracy by at most $0.4$ points from epoch 200 for the four-layer heads and leaves \qpmel{} at $19.9\%$ ($25.7\%$ train at epoch 200). On QNervFormer, \qpmel{} ends at a training loss of $0.62$ and one-layer DR between $1.6$ and $4.7$, while the four-layer circuit heads end between $5.9$ and $6.9$, near $\ln 1024 = 6.93$, the loss of a uniform softmax over the batch. Under the $\langle Z \rangle$ cosine readout of \citet{chen2024quantum}, DR fits less than under fidelity: $0.8 \pm 0.4\%$ of QNervFormer's training images and $0.5 \pm 0.6\%$ of QSTConv's at four layers, and at most $3.3\%$ at any depth. Our E-VQC matches the equations of \citet{chen2024quantum} to $1.8 \times 10^{-7}$ but reaches $0.84\%$ held-out on QSTConv against their $3.5\%$ average over subjects; the paper has no public code, leaves seven architectural choices unspecified and does not fully specify its evaluation protocol.

\paragraph{Regularization.} The branch-dropout columns of Table~\ref{tab:eeg} add dropout $0.5$ at the end of each convolution branch of QNervFormer, and Table~\ref{tab:dr_depth} applies the same protocol to DR at every depth. \qpmel{}'s $11.8 \pm 5.6\%$ spread there comes from one seed that does not take off within 200 epochs; of the eight runs of this configuration, seven reach $12.0$--$17.5\%$ held-out (median $15.0\%$). Attention dropout, feed-forward dropout or per-sample normalization alone leaves memorization in place ($86$--$98\%$ train at epoch 200). Holding \qpmel{}'s kernel additive for all of training does not prevent memorization (Table~\ref{tab:curriculum}), and fixing the logit scale at 5 slows it to $87.7\%$ train without raising held-out accuracy.

\paragraph{Curriculum ablation.} Table~\ref{tab:curriculum} holds \qpmel{}'s kernel additive for all 200 epochs, so it never becomes the product \pmef{}, and compares it with the additive-to-product curriculum of Eq.~\ref{eq:curriculum} over the full grid of 10 subjects and 5 seeds. On QSTConv the product end of the curriculum adds about 5 points of top-1. On QNervFormer both schedules memorize: the additive kernel's mean unmatched fidelity over a batch of $B$ is at least $\tfrac{1}{2}\bigl(1 - 1/(B-1)\bigr)$, so it cannot place the batch at orthogonal states, yet its training loss still falls to $1.40$ with held-out accuracy at chance.

\begin{table}[h]
\centering
\footnotesize
\setlength{\tabcolsep}{3pt}
\caption{\textit{\qpmel{} kernel schedule on THINGS-EEG2, 10 subjects and 5 seeds: 200-way held-out accuracy (\%), final training loss, and runs whose final loss stays above $6.9$ of $50$. Here, Curr. is Curriculum and Add. is Additive.}}
\label{tab:curriculum}
\begin{tabular}{lcccccccc}
\toprule
& \multicolumn{2}{c}{Top-1} & \multicolumn{2}{c}{Top-5} & \multicolumn{2}{c}{Final loss} & \multicolumn{2}{c}{Collapsed} \\
\cmidrule(lr){2-3} \cmidrule(lr){4-5} \cmidrule(lr){6-7} \cmidrule(lr){8-9}
Encoder & Curr. & Add. & Curr. & Add. & Curr. & Add. & Curr. & Add. \\
\midrule
QSTConv & $\mathbf{23.96 \pm 5.2}$ & $19.11 \pm 3.9$ & $\mathbf{56.82 \pm 7.6}$ & $48.56 \pm 5.5$ & $4.81$ & $5.38$ & $0$ & $0$ \\
QNervFormer & $0.92 \pm 0.8$ & $0.73 \pm 0.5$ & $4.92 \pm 2.0$ & $3.93 \pm 1.4$ & $0.50$ & $1.40$ & $0$ & $0$ \\
\bottomrule
\end{tabular}
\end{table}

\subsection{Kernel and gradient diagnostics} \label{app:trainability}
Table~\ref{tab:trainability} collects the measurements behind Section~\ref{subsec:trainability_results} on QNervFormer: kernel statistics at initialization on 10 qubits with four circuit layers and inputs drawn uniformly from $[-\pi,\pi]$, the RMS gradient each head passes to the encoder at the first epoch, and the fit to training images after 200 epochs. Range is the spread of kernel values times CLIP's initial logit scale of $14.29$, the logit separation available to the softmax, and structure is the rank correlation $\rho$ of the kernel with input similarity; \qpmel{}'s structure was not measured (n/m). Each failing kernel lacks one of the two. Under fidelity the E-VQC's trainable layers cancel (Eq.~\ref{eq:terminal_isometry}), leaving the single-axis $R_Y$ product kernel, structured ($\rho = 0.79$) but concentrated at $2^{-10}$ and worth $0.09$ nats. The $\langle Z \rangle$ cosine readout has range ($6.6$ nats) but little structure ($\rho = 0.19$). Concentration does not require entanglement: \qpmel{}'s product kernel also starts at $2^{-10}$, and its additive kernel is what gives it $1.39$ nats. No failing head has a vanishing gradient. The $\langle Z \rangle$ E-VQC passes the encoder 17 times \qpmel{}'s gradient at the first epoch and fits 30 times worse, and scoring the E-VQC by the log of its fidelity gives the largest gradient ($955$) and the worst fit ($0.3\%$). The E-VQC's circuit-gradient variance $\Var[\partial \langle Z_0 \rangle / \partial w]$ falls from $0.22$ to $2.3 \times 10^{-3}$ between 2 and 12 qubits, and its per-parameter gradient in mid-training shows no depth trend from 1 to 16 layers that survives a rerun. Redrawing the E-VQC's trainable weights changes its fidelity kernel by at most $1.6 \times 10^{-6}$ of its mean, as Eq.~\ref{eq:terminal_isometry} requires. Its entangling layers shrink the single-qubit readout, with mean $|\langle Z_i \rangle|$ falling from $0.27$ at one layer to $0.10$ at four, and cosine renormalizes what remains.

\begin{table}[t]
    \centering
    \footnotesize
    \setlength{\tabcolsep}{4pt}
    \caption{\textit{Kernel and gradient diagnostics on THINGS-EEG2 with QNervFormer; four-layer circuits.}}
    \label{tab:trainability}
    \begin{tabular}{llccccc}
        \toprule
        & & \multicolumn{3}{c}{Kernel at initialization} & & \\
        \cmidrule(lr){3-5}
        Head & Similarity & Mean & Range (nats) & Structure $\rho$ & Encoder gradient & Train (\%) \\
        \midrule
        E-VQC & $\langle Z \rangle$ cosine & $-3.9 \times 10^{-3}$ & $6.57$ & $0.19$ & $10.6$ & $3.2$ \\
        E-VQC & Fidelity & $8.8 \times 10^{-4}$ & $0.09$ & $0.79$ & $0.010$ & $10.1$ \\
        E-VQC & $\log$ fidelity & $8.8 \times 10^{-4}$ & -- & $0.79$ & $955$ & $0.3$ \\
        DR & Fidelity & $1.0 \times 10^{-3}$ & $0.02$ & $0.06$ & $0.021$ & $6.5$ \\
        \midrule
        \multirow{2}{*}{\qpmel{}} & \pmef{}, product & $9.9 \times 10^{-4}$ & $0.07$ & n/m & \multirow{2}{*}{$0.62$} & \multirow{2}{*}{$\mathbf{96.5}$} \\
        & \pmef{}, additive & $0.50$ & $1.39$ & n/m & & \\
        \bottomrule
    \end{tabular}
\end{table}

\paragraph{Re-uploading depth.} Table~\ref{tab:dr_depth} varies the number of layers $L$ of the two circuits scored by fidelity. Each DR layer applies a trainable $R_Y(w_q)$ and a data $R_Z(x_q)$ to every qubit, then a CNOT ring. The ring after the last upload is the same unitary in both towers and cancels from the fidelity (Eq.~\ref{eq:terminal_isometry}), so the kernel contains $L-1$ rings. At $L = 1$ it is the product kernel $\prod_q |\bra{0} R_Y(-w_q)\, R_Z(x'_q - x_q)\, R_Y(w'_q) \ket{0}|^2$ on $\MU$, with primes marking the image tower, in which the encoder sets only the azimuth of each qubit and the polar angle is a trained constant of each tower. When the two towers' angles agree, qubit $q$ contributes $1 - \sin^2 w_q \sin^2((x_q - x'_q)/2)$, so $w_q$ sets how strongly the qubit separates inputs, from not at all at $w_q = 0$ to the E-VQC's $\cos^2((x_q - x'_q)/2)$ at $w_q = \pi/2$. On the memorizing QNervFormer this is the best-fitting circuit head in the study ($65.0 \pm 17.9\%$ over ten seeds, range $40.0$--$85.5\%$), and at one and two layers DR fits more than the E-VQC, whose trainable layers cancel. Each added layer removes structure from DR's kernel, and fit falls with it; by $L = 8$ the kernel's mean and spread match those of Haar-random states. The E-VQC's kernel is the same at every depth within a tower, and its fit settles near $10\%$ from four layers on. No run on this backbone exceeds $3.0\%$ held-out.

On the two backbones that generalize, held-out accuracy falls with depth as fit does. One-layer DR reaches $12.3 \pm 5.2\%$ held-out on QSTConv and $6.6 \pm 4.6\%$ on QNervFormer with branch dropout, against $19.9 \pm 3.1\%$ and $11.8 \pm 5.6\%$ for \qpmel{} under the same selection; its best seed on QSTConv reaches $16.5\%$, but its mean trails \qpmel{} on both. Two layers, which leave one ring in the kernel, give $3.7\%$ and $2.3\%$, and four or more stay within half a point of chance. The circuit head that generalizes best is therefore the one whose entanglement cancels, and it trails \qpmel{}, which differs from it both in setting both angles of every qubit and in beginning training with the additive kernel; these factors are not separated here.

The first-epoch encoder gradient rises with depth, so the collapse is not a vanishing gradient at initialization. The one-layer circuit passes the smallest gradient because its towers' polar angles start as independent draws: the kernel between towers then starts concentrated, averaging $3.6 \times 10^{-4}$ at initialization against $2^{-10} \approx 9.8 \times 10^{-4}$, and its gradient is multiplicative (Eq.~\ref{eq:mult_gradient}). Every one-layer run starts on the $\ln 1024$ plateau, and when it leaves sets most of the spread. On QNervFormer runs leave between epochs 30 and 90, and the epoch of departure has rank correlation $-0.81$ with the final fit over ten seeds. With branch dropout they leave later, between epochs 75 and 125, and the one seed that never leaves ends at $0.5\%$ held-out. On QSTConv all five seeds leave by epoch 100, but two climb more slowly and end at $5.5$ and $8.0\%$ against $15.5$--$16.5\%$ for the rest. On these two backbones deeper circuits leave later or not at all: two-layer runs leave between epochs 75 and 150, and no four-layer run on QSTConv leaves. \qpmel{}'s loss instead falls from the first epoch on both regularized backbones, as the range of its additive kernel allows (Table~\ref{tab:trainability}). Circuit gradients measured at epoch 100 do fall with depth, by $40\times$ from one layer to sixteen, but by then the shallow runs have left the plateau and the deep ones have not. At four layers the separate five-seed runs of Table~\ref{tab:eeg} give $6.5 \pm 1.7\%$ train.

\begin{table}[h]
    \centering
    \footnotesize
    \setlength{\tabcolsep}{4pt}
    \caption{\textit{Circuit depth under the fidelity readout on THINGS-EEG2. Structure is measured at initialization within one tower on inputs uniform in $[-\pi,\pi]$, and the encoder gradient is the RMS at the first epoch on QNervFormer. Train is the fit to 200 training images on QNervFormer after 200 epochs, three seeds ($^\ddagger$ten). Held-out rows are on QSTConv and on QNervFormer with branch dropout, with the checkpoint selection of Table~\ref{tab:eeg}, five seeds ($^\dagger$four); \qpmel{} reaches $19.9$ and $11.8\%$ there. n/r, not run.}}
    \label{tab:dr_depth}
    \begin{tabular}{llccccc}
        \toprule
        & & \multicolumn{5}{c}{Layers $L$} \\
        \cmidrule(lr){3-7}
        Head & & 1 & 2 & 4 & 8 & 16 \\
        \midrule
        \multirow{6}{*}{DR} & Rings in the kernel & $0$ & $1$ & $3$ & $7$ & $15$ \\
        & Structure $\rho$ & $0.64$ & $0.46$ & $0.06$ & $0.00$ & $0.00$ \\
        & Encoder gradient & $1.4 \times 10^{-4}$ & $8.3 \times 10^{-3}$ & $0.022$ & $0.041$ & $0.068$ \\
        & Train (\%) & $\mathbf{65.0 \pm 17.9}^\ddagger$ & $43.0 \pm 8.0$ & $7.0 \pm 3.1$ & $1.0 \pm 0.0$ & $2.0 \pm 0.9$ \\
        & Held-out, QSTConv & $\mathbf{12.3 \pm 5.2}$ & $3.7 \pm 1.2$ & $0.7 \pm 0.4$ & n/r & n/r \\
        & Held-out, dropout & $\mathbf{6.6 \pm 4.6}$ & $2.3 \pm 2.0$ & $0.9 \pm 0.7$ & $0.6 \pm 0.2$ & $0.2 \pm 0.5^\dagger$ \\
        \midrule
        \multirow{2}{*}{E-VQC} & Structure $\rho$ & $0.80$ & $0.80$ & $0.79$ & $0.79$ & $0.79$ \\
        & Train (\%) & $25.8 \pm 2.3$ & $23.2 \pm 1.5$ & $12.8 \pm 9.1$ & $9.8 \pm 2.5$ & $8.8 \pm 2.0$ \\
        \bottomrule
    \end{tabular}
\end{table}

\paragraph{Trained kernels.} Table~\ref{tab:kernel_geometry} measures the kernel each run ends with, on the first 1024 training pairs at the checkpoint Table~\ref{tab:eeg} reports. For Haar-random states on 10 qubits both kinds of pair average $2^{-10}$, and a fraction $1 - (1 - 0.01)^{1023} > 0.9999$ of unmatched pairs falls below $0.01$. With depth, DR's kernel moves to these values on both backbones: the gap between matched and unmatched pairs closes and the fraction of unmatched pairs below $0.01$ rises, reaching the Haar value at four layers on QSTConv and at sixteen with branch dropout. \qpmel{} ends with the largest gap of any generalizing run on both backbones, $0.29$ on QSTConv against $0.12$ for one-layer DR, with only $0.2\%$ of unmatched pairs below $0.01$. Unregularized QNervFormer under \qpmel{} instead sends a third of unmatched pairs below $0.01$ and lowers their mean to $0.05$, with its learned logit scale at $24$. Fixing the scale at 5 leaves $45\%$ of them below $0.01$ at epoch 200, and holding the kernel additive, which bounds the mean unmatched fidelity of a batch below by nearly $\tfrac{1}{2}$, still memorizes. The memorizing encoder therefore reaches the shattered configuration where the product kernel allows it, but does not need it to memorize.

\paragraph{Backbones.} The two encoders order the heads identically and differ in how far they move each kernel. Without regularization QNervFormer fits every head more than QSTConv, and with branch dropout it still holds deep kernels away from the Haar value: at four layers its unmatched pairs average $0.029$, 30 times $2^{-10}$, and $60\%$ fall below $0.01$, against $1.2 \times 10^{-3}$ and $99\%$ on QSTConv. The encoder gradients follow. Under four-layer DR on QNervFormer the RMS encoder gradient grows from $0.021$ at the first epoch to $0.17$ at epoch 100 and $0.64$ at epoch 200, and runs spend $3.4$ of 9 checkpoints on the plateau. On QSTConv it decays from $0.042$ to $0.022$ and $0.019$, and every checkpoint of every run sits on the plateau. The same flexibility lets unregularized QNervFormer shatter under \qpmel{}, and dropout on its convolution branches returns it to QSTConv's regime: its \qpmel{} kernel ($0.50$ matched, $0.25$ unmatched, $0.2\%$ below $0.01$) then resembles QSTConv's ($0.62$, $0.34$, $0.2\%$) rather than its own unregularized one.

\begin{table}[h]
    \centering
    \footnotesize
    \setlength{\tabcolsep}{3pt}
    \caption{\textit{Trained kernels on THINGS-EEG2: mean fidelity of matched and of unmatched pairs, and the fraction of unmatched pairs below $0.01$, on 1024 training pairs at the checkpoint of Table~\ref{tab:eeg}; five seeds ($^\dagger$four). $^\ddagger$QNervFormer without branch dropout at epoch 200, three seeds. Haar-random states give $9.8 \times 10^{-4}$, $9.8 \times 10^{-4}$ and $1.00$. n/r, not run.}}
    \label{tab:kernel_geometry}
    \begin{tabular}{llcccccc}
        \toprule
        & & \multicolumn{3}{c}{QSTConv} & \multicolumn{3}{c}{QNervFormer, branch dropout} \\
        \cmidrule(lr){3-5} \cmidrule(lr){6-8}
        Head & $L$ & Matched & Unmatched & $<0.01$ & Matched & Unmatched & $<0.01$ \\
        \midrule
        E-VQC, fidelity & 4 & $0.057$ & $0.034$ & $0.39$ & $0.127$ & $0.072$ & $0.18$ \\
        DR & 1 & $0.282$ & $0.161$ & $0.02$ & $0.214$ & $0.121$ & $0.22$ \\
        DR & 2 & $0.188$ & $0.117$ & $0.06$ & $0.123$ & $0.072$ & $0.28$ \\
        DR & 4 & $1.3 \times 10^{-3}$ & $1.2 \times 10^{-3}$ & $0.99$ & $0.047$ & $0.029$ & $0.60$ \\
        DR & 8 & n/r & n/r & n/r & $0.019$ & $8.9 \times 10^{-3}$ & $0.87$ \\
        DR & $16^\dagger$ & n/r & n/r & n/r & $9.7 \times 10^{-4}$ & $9.7 \times 10^{-4}$ & $1.00$ \\
        \qpmel{} & -- & $0.625$ & $0.336$ & $0.002$ & $0.501$ & $0.252$ & $0.002$ \\
        \midrule
        \qpmel{}, no dropout$^\ddagger$ & -- & -- & -- & -- & $0.557$ & $0.053$ & $0.32$ \\
        \bottomrule
    \end{tabular}
\end{table}

\subsection{MS-COCO} \label{app:coco}
MS-COCO Captions \citep{chen2015coco} runs \qpmel{} at 20 qubits with pretrained Xception \citep{chollet2017xception} and BERT \citep{devlin2019bert} encoders, trained with the symmetric InfoNCE objective on $30{,}000$ training images and their captions. Table~\ref{tab:coco_full} reports cross-modal matching on training pairs in both directions, for the closed-form \pmef{} and for statevector simulation of the 20-qubit circuit, with confidence intervals over episodes. \qpmel{} matches an image to its caption among 15 candidates with $94.7\%$ accuracy and a caption to its image with $95.9\%$ (chance $6.7\%$), and the two kernels agree within 3 points at every way count. The accuracies are measured on training pairs, so they show fit at 20 qubits, not generalization.

\begin{table}[h]
    \centering
    \footnotesize
    \setlength{\tabcolsep}{5pt}
    \caption{\textit{MS-COCO cross-modal matching on training pairs (\%), 20 qubits: closed-form \pmef{} and statevector simulation of the same circuit. Chance is $100/n$ for $n$-way.}}
    \label{tab:coco_full}
    \begin{tabular}{lcccc}
        \toprule
        & \multicolumn{2}{c}{Image $\rightarrow$ text} & \multicolumn{2}{c}{Text $\rightarrow$ image} \\
        \cmidrule(lr){2-3} \cmidrule(lr){4-5}
        Way & \pmef{} & Simulated & \pmef{} & Simulated \\
        \midrule
        3 & $99.6 \pm 0.6$ & $98.7 \pm 1.9$ & $98.9 \pm 1.0$ & $98.0 \pm 2.3$ \\
        5 & $98.5 \pm 0.9$ & $98.8 \pm 1.4$ & $98.5 \pm 0.8$ & $98.0 \pm 1.7$ \\
        7 & $97.8 \pm 0.9$ & $96.3 \pm 2.1$ & $97.5 \pm 1.1$ & $96.3 \pm 2.4$ \\
        9 & $98.1 \pm 0.8$ & $95.3 \pm 2.1$ & $97.0 \pm 1.0$ & $95.1 \pm 2.4$ \\
        11 & $96.5 \pm 0.9$ & $94.4 \pm 2.1$ & $96.8 \pm 0.8$ & $95.1 \pm 1.7$ \\
        13 & $95.8 \pm 0.9$ & $95.2 \pm 1.8$ & $95.6 \pm 0.9$ & $93.7 \pm 2.2$ \\
        15 & $94.7 \pm 1.0$ & $93.6 \pm 1.9$ & $95.9 \pm 0.9$ & $93.1 \pm 1.7$ \\
        \bottomrule
    \end{tabular}
\end{table}

\subsection{Simulated and Executed Circuits: Inversion vs.\ SWAP Test} \label{appendix:circuit_discrepancy}

The noiseless and noisy simulations do not run the circuit that the physical QPU runs. Both estimate the same quantity, the product-state fidelity $|\braket{\psi(x)|\psi(y)}|^2$, but through two different primitives with very different hardware cost. We state the discrepancy explicitly because it determines how the hardware numbers in Table~\ref{tab:hardware} should be read.

\paragraph{Simulation: the inversion test.}
The simulator prepares $\ket{\psi(x)}$ on $N$ qubits and then applies the inverse preparation for $\ket{\psi(y)}$, measuring the probability of returning to the all-zeros string, $P(\ket{0}^{\otimes N}) = |\braket{\psi(y)|\psi(x)}|^2$. Because both states are products, the circuit is $R_Y(2\theta^x_q)$, $R_Z(\gamma^x_q)$, $R_Z(-\gamma^y_q)$, $R_Y(-2\theta^y_q)$ on each qubit independently: width $N$, depth $4$, and \emph{zero} two-qubit gates. Transpiled against the IBM heavy-hex basis this remains depth $6$ with no entangling gates at every qubit count tested.

\paragraph{Hardware: the SWAP test.}
The QPU runs a SWAP test instead. It uses $2N+1$ qubits, one control plus two $N$-qubit registers holding $\ket{\psi(x)}$ and $\ket{\psi(y)}$, applies a Hadamard on the control, $N$ controlled-SWAP gates between corresponding register qubits, a second Hadamard, and measures only the control. The fidelity is recovered as $2P(0)-1$. The historical implementation used this primitive with \texttt{SamplerV2} parameter binding and one-bit readout.

\paragraph{Cost of that choice.}
The SWAP test is not free. Each CSWAP decomposes into $8$ CNOT-equivalent two-qubit gates, so the logical circuit carries $8N$ entangling gates where the inversion test carries none, and heavy-hex routing adds more. Table~\ref{appendix:circuit_cost} reports transpiled counts for both primitives against the \texttt{ibm\_marrakesh} coupling map at optimization level $1$.

\begin{table}[h!]
    \centering
    \footnotesize
    \renewcommand{\arraystretch}{1.2}
    \caption{\textit{Transpiled cost of the two fidelity primitives, targeting the \texttt{ibm\_marrakesh} coupling map and basis at optimization level $1$. Two-qubit counts are native \texttt{cz} gates after routing; the pre-routing logical count is $8N$.}}
    \label{appendix:circuit_cost}
    \begin{tabular}{llrrr}
        \toprule
        \textbf{Primitive} & \textbf{Setting} & \textbf{Qubits} & \textbf{2-qubit gates} & \textbf{Depth} \\
        \midrule
        Inversion test & $N=5$  & 5  & 0   & 6 \\
        Inversion test & $N=6$  & 6  & 0   & 6 \\
        Inversion test & $N=10$ & 10 & 0   & 6 \\
        \midrule
        SWAP test & $N=5$  & 11 & 73  & 187 \\
        SWAP test & $N=6$  & 13 & 93  & 232 \\
        SWAP test & $N=10$ & 21 & 167 & 394 \\
        \bottomrule
    \end{tabular}
\end{table}

At $N=10$ the recorded transpilation uses roughly twice the qubits, $167$ two-qubit gates instead of none, and about $65\times$ the depth of the inversion circuit. The comparison between the $3000$-shot simulation and the $128$-shot hardware run confounds shot noise, device noise, and the estimator circuit.